\documentclass[amsmath, pra, twocolumn, showpacs]{revtex4-1}
\usepackage{graphicx}
\usepackage{dsfont}

\begin{document}     

\title{Anisotropy in scattering of light from an atom into the guided modes of a nanofiber}
 
\author{Fam Le Kien}

\author{A. Rauschenbeutel} 

\affiliation{Vienna Center for Quantum Science and Technology, Institute of Atomic and Subatomic Physics, Vienna University of Technology, Stadionallee 2, 1020 Vienna, Austria}

\date{\today}

\begin{abstract}
We study the scattering of guided light from a multilevel cesium atom with the transitions between the hyperfine levels $6S_{1/2}F=4$ and $6P_{3/2}F'=5$ of the $D_2$ line into the guided modes of a nanofiber.
We show that the rate of scattering of guided light from the atom in the steady-state regime into the guided modes 
is asymmetric with respect to the forward and backward directions and depends on the polarization of the probe field. The asymmetry between the forward and backward scattering is a result of the complex transition structure of the atom and the existence of a longitudinal component of the guided-mode profile function. In the case of a two-level atom, the rates of spontaneous emission (and consequently the rates
of scattering) into the forward and backward guided modes differ from each other when the atomic dipole matrix-element vector
is a complex vector in the plane that contains the fiber axis and the atomic position. 
\end{abstract}

\pacs{42.50.Nn, 42.50.Ct, 42.81.Dp, 42.81.Gs}
\maketitle

\section{Introduction}
\label{sec:introduction}

Over the last decade, optical fibers that are tapered to a diameter comparable to or smaller than the wavelength of light \cite{Mazur's Nature,Birks,taper}
have attracted considerable attention for a wide range of potential practical applications \cite{Morrissey13}.
In such a thin fiber, called a nanofiber, the guided field penetrates an appreciable distance into the surrounding medium and appears as an evanescent wave carrying a significant fraction of the propagation power and having a complex polarization pattern \cite{Bures99,Tong04,fibermode}. Nanofiber-guided light fields can be used for trapping atoms \cite{fiber trap,Vetsch10,Goban12}, for probing atoms \cite{Domokos02,absorption,Nayak07,Nayak09,Dawkins11,Reitz13,Russell13}, molecules \cite{Stiebeiner09}, quantum dots \cite{Yalla12}, and color centers in nanodiamonds \cite{Schroder12,Liebermeister13}, and for mechanical manipulations of small particles \cite{Skelton12,Brambilla07,Fam13}. 

The ability to control and manipulate atoms individually is of great importance for various applications in both fundamental and applied physics \cite{Schlosser,Kuhr,Sackett}.
In order to find an effective way to probe, control, and manipulate an atom trapped outside a nanofiber, we need to know the optical response of the atom to a near-resonant field propagating along the fiber. 
Absorption and scattering are the usual outcomes of the interaction of an atom with a near-resonant light field. 
The absorption and scattering of guided light by a single atom have been studied  \cite{Domokos02,absorption}.
It has been shown by Domokos \textit{et al.} \cite{Domokos02} for a two-level atom that, due to the transverse confinement of the field in a waveguide, a single atom is able to have a significant effect on a wave packet of light. When the transverse extension of the field in a guided mode is close to the radiative cross section of the atom, the latter becomes a significant scatterer. Similar to the radiation of an oscillating electric dipole, the scattering of light from a two-level atom with a real dipole matrix-element vector in free space has equal rates for the forward and backward directions \cite{Loudon,Scully,Mandel}. This property is a consequence of the point symmetry of the system. The effect of the multilevel structure of a real atom on the absorption and scattering characteristics has been examined \cite{absorption}. While the general formalism and the results of calculations for the total scattering rate in Ref.~\cite{absorption} are correct, the phenomenological separation of the total scattering rate into two equal components for backward and forward scattering is naive and generally not correct. For an atom with a multilevel structure or with a complex dipole matrix-element vector in the vicinity of an object, the point symmetry may be broken. Recent experimental progress has demonstrated that the scattering of guided light from  realistic multilevel atoms is very different from the case of atoms in free space \cite{Reitz14,Mitsch14}. Therefore, it is necessary to develop a systematic microscopic theory for the forward and backward scattering of guided light from a multilevel atom taking into account the complexity of the atomic dipole polarization and the field polarization. 

Before we proceed, we note that the excitation of a multilevel atom by laser light of arbitrary polarization
has been thoroughly studied \cite{Milner98,Milner99,Taichenachev99,Vitanov03,Taichenachev04,Yudin13,ChangMinogin}. It has been shown that elliptically polarized light creates an anisotropic distribution
of atomic angular momentum \cite{Milner99}. Scattering of guided light from an atom involves not only the atomic excitation but also the subsequent spontaneous emission. 
The latter is a quantum electrodynamic process caused by vacuum fluctuations. The presence of the fiber opens the channel of spontaneous emission into guided modes, modifies the rate of spontaneous emission into radiation modes, and leads to the appearance of cross-level decay coefficients \cite{cesium decay}.

In this paper, we study the scattering of guided light from a multilevel atom into  the forward and backward guided modes of a nanofiber. We show that the scattering rate is asymmetric with respect to the forward and backward directions and depends on the polarization of the probe field.
   
The paper is organized as follows. In Sec.\ \ref{sec:theory} we study the scattering of guided light from a multilevel atom. In Sec.\ \ref{sec:twolevelatom} we discuss the directional spontaneous emission of a two-level atom with a complex dipole matrix-element vector. Our conclusions are given in Sec.~\ref{sec:summary}.

\section{Scattering of guided light from a multilevel atom}
\label{sec:theory}

\begin{figure}[tbh]
\begin{center}
  \includegraphics{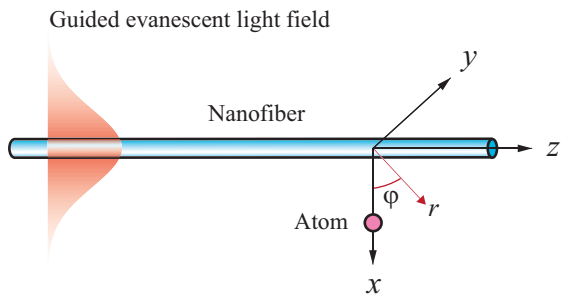}
 \end{center}
\caption{(Color online) 
Probing an atom by an evanescent light field propagating along a thin optical fiber.}
\label{fig1}
\end{figure}

Consider the scattering of a guided light field from a single alkali-metal atom trapped outside an optical nanofiber 
(see Fig.~\ref{fig1}). The nanofiber has a cylindrical silica core, with the radius $a$ and the refractive index $n_1=1.45$, surrounded by vacuum, with the refractive index $n_2=1$. 
The diameter $2a$ of the nanofiber is comparable to or smaller than the wavelength $\lambda$ of light. 
Such a thin fiber can be produced by the taper fiber technology  \cite{Mazur's Nature,Birks,taper}. 
The essence of the technology is to heat and pull a single-mode optical fiber to a very small thickness, maintaining the taper condition to keep adiabatically the single-mode condition. Due to tapering, the original core is almost vanishing. Therefore, the refractive indices that determine the guiding properties of the tapered fiber are the refractive index of the original silica clad and the refractive index of the surrounding vacuum. Subwavelength-diameter vacuum-clad silica-core fibers are nanofibers. 

In view of the very low losses of silica in the wavelength range of interest, we neglect material absorption. 
In the presence of the fiber, the electromagnetic field can be decomposed into guided modes and radiation modes \cite{fiber books}. 
The guided modes have the evanescent behavior on the outside of the core. They can travel in the waveguide without loss of power, provided that losses in the dielectric material are ignored. Meanwhile, the radiation modes are oscillatory at large distances from the fiber and do not have the evanescent behavior. They cannot be normalized to a finite amount of power.

We assume that the single-mode condition \cite{fiber books} is satisfied for a finite bandwidth around a central atomic transition frequency $\omega_0$. 
Although our theory is general and applicable, in principle, to an arbitrary multilevel atom, we assume a cesium atom throughout this paper. For simplicity, 
we neglect the effect of the surface-induced potential on the atomic energy levels. This approximation is good when the atom is not too close to the fiber surface \cite{Fam07}.

\subsection{Interaction of the atom with the guided field}

We use the Cartesian coordinates  $\{x,y,z\}$ and the associated cylindrical coordinates $\{r,\varphi,z\}$, with $z$ being the fiber axis (see Fig.~\ref{fig1}). 
We represent the electric component of the guided light field as
$\mathbf{E}=(\boldsymbol{\mathcal{E}}e^{-i\omega t}+\mathrm{c.c.})/2
=(\mathcal{E}\mathbf{u}e^{-i\omega t}+\mathrm{c.c.})/2$,
where $\omega$ is the angular frequency and $\boldsymbol{\mathcal{E}}=\mathcal{E}\mathbf{u}$ is the slowly varying envelope 
of the positive-frequency part,
with $\mathcal{E}$ and $\mathbf{u}$ being the field amplitude and the polarization vector, respectively.
We assume that the guided probe field $\mathbf{E}$ propagates in the positive direction $+z$, from the left-hand side to the right-hand side of Fig.~\ref{fig1}.
In general, the amplitude $\mathcal{E}$ is a complex scalar and the polarization vector $\mathbf{u}$ is a complex unit vector. 
The guided light field can be decomposed into a superposition of quasicircularly or quasilinearly polarized modes \cite{fiber books}.
In the cylindrical coordinates, the electric component of the guided field 
is given, in the case of quasicircularly polarized modes, by
\begin{equation}\label{x1}
\boldsymbol{\mathcal{E}}_{\mathrm{circ}} 
= A_{\mathrm{circ}}(\hat{\mathbf{r}}e_r+l\hat{\boldsymbol{\varphi}}e_\varphi+
f\hat{\mathbf{z}}e_z) e^{if\beta z +il\varphi}
\end{equation}
and, in the case of quasilinearly polarized modes, by
\begin{eqnarray}\label{x2}
\boldsymbol{\mathcal{E}}_{\mathrm{lin}}
&=&A_{\mathrm{lin}}[\hat{\mathbf{r}}e_r\cos(\varphi-\varphi_0)
+i\hat{\boldsymbol{\varphi}}e_\varphi\sin(\varphi-\varphi_0)
\nonumber\\&&\mbox{}
+f\hat{\mathbf{z}}e_z\cos(\varphi-\varphi_0)] 
e^{if\beta z}.
\end{eqnarray}
Here the notations $\hat{\mathbf{r}}=\hat{\mathbf{x}}\cos\varphi + \hat{\mathbf{y}}\sin\varphi$, 
$\hat{\boldsymbol{\varphi}}=-\hat{\mathbf{x}}\sin\varphi + \hat{\mathbf{y}}\cos\varphi$, and $\hat{\mathbf{z}}$
stand for the unit basis vectors of the cylindrical coordinate system, where $\hat{\mathbf{x}}$ and $\hat{\mathbf{y}}$ are the unit basis vectors of the Cartesian coordinate system for the fiber transverse plane $xy$. The index $f=+1$ or $-1$ (or simply $+$ or $-$) stands for the forward ($+\hat{\mathbf{z}}$) or backward ($-\hat{\mathbf{z}}$) propagation direction, respectively, and
the index $l=+1$ or $-1$ (or simply $+$ or $-$) refers to the counterclockwise or
clockwise circulation, respectively, of the transverse component of the field with respect to the positive direction of the fiber axis $z$. 
The angle $\varphi_0=0$ or $\pi/2$ in Eq.~(\ref{x2}) for quasilinearly polarized modes specifies the principal direction $x$ or $y$, respectively, of the polarization vector $\mathbf{u}$ in the fiber transverse plane $xy$. The parameter $\beta$ is the longitudinal propagation constant for the fiber fundamental mode. 
The explicit expressions for the cylindrical components $e_r(r)$, $e_{\varphi}(r)$, and $e_z(r)$ of the guided-mode profile function $\mathbf{e}(r,\varphi,z)$ 
are given in Refs.~\cite{fiber books,fibermode,cesium decay} and are summarized in Appendix \ref{sec:guided}. The coefficients $A_{\mathrm{circ}}$ and $A_{\mathrm{lin}}$
can be determined from the propagation power $P_z$ of the guided light field. 
The power $P_z$ is given by the formula \cite{fiber books}
\begin{equation}\label{x2a}
P_z=\frac{\epsilon_0v_g}{2}\int n_{\mathrm{ref}}^2(r)|\boldsymbol{\mathcal{E}}(\mathbf{r})|^2 \,d^2\mathbf{r},
\end{equation}
where $v_g=1/\beta'(\omega)\equiv(d\beta/d\omega)^{-1}$ is the group velocity of guided light,
$n_{\mathrm{ref}}(r)=n_1$ and $n_2$ for $r<a$ and $r>a$, respectively, is the position-dependent refractive index,  
and $\int d^2\mathbf{r}=\int_0^{2\pi}d\varphi\int_0^{\infty}r\,dr$ is the integral over the fiber cross-section plane.
The notation $\beta'$ stands for the derivative of the propagation constant $\beta$ with respect to the frequency $\omega$. 
It is interesting to note from Eqs.~\eqref{x1} and \eqref{x2} that the difference between the forward and backward guided fields is expressed by
not only the change in sign of the phase factor $f\beta z$ but also
the change in sign of the longitudinal component $fe_z$ (see also Appendix \ref{sec:guided}).
The latter may affect the magnitude of the coupling between the atom and the field and, consequently, may cause a difference between the forward and backward scattering.
However, as will be shown later, the existence of a longitudinal component of the guided field is just a necessary condition but not an enough condition for asymmetry between the forward and backward scattering.  

We study the $D_2$ line of atomic cesium, which  occurs at the wavelength $\lambda_0=852$ nm and 
corresponds to the transition from the ground state $6S_{1/2}$
to the excited state $6P_{3/2}$. We assume that the cesium atom is initially prepared in the hyperfine-structure (hfs) level $F=4$ of the ground state $6S_{1/2}$ and that the probe field is tuned close to resonance with the transition 
from this ground-state hfs level to the hfs level $F'=5$ of the excited state  $6P_{3/2}$.
Among the hfs components  of the $D_2$ line,  
the transition $6S_{1/2}F=4\leftrightarrow 6P_{3/2}F'=5$ has the strongest oscillator strength.
Because of the selection rule
$\Delta F=0,\pm1$, spontaneous emission from the excited hfs level $6P_{3/2}F'=5$ to the ground state is always to the ground-state hfs level $6S_{1/2}F=4$, not
to the other ground-state  hfs level $6S_{1/2}F=3$. Therefore, the magnetic (Zeeman) sublevels of the hfs levels $6S_{1/2}F=4$ and $6P_{3/2}F'=5$ form a closed set, which is used for laser cooling in magneto-optical traps \cite{coolingbook}.

In order to describe the internal state of the cesium atom, we use the fiber axis $z$ as the quantization axis.
In addition, we assume that the atom is located on the positive side of the axis $x$ unless stated otherwise \cite{Mitsch14}. For convenience, we introduce the notations 
$|e\rangle\equiv|F'M'\rangle$ and $|g\rangle\equiv|FM\rangle$ for the magnetic sublevels 
$F'M'$ and $FM$ of the hfs levels $6P_{3/2}F'=5 $ and $6S_{1/2}F=4$, 
respectively. The $q$ spherical tensor component $d_{M'M}^{(q)}$ of the  dipole matrix-element vector $\mathbf{d}_{M'M}$  for the transition between $|F'M'\rangle$ and $|FM\rangle$, where $q=M'-M=0,\pm1$, is given by the formula \cite{Shore}
\begin{eqnarray}\label{x3}
d_{M'M}^{(q)}&=&(-1)^{I+J'-M'}\langle J' \| D\| J\rangle\sqrt{(2F+1)(2F'+1)}\nonumber\\
&&\mbox{}\times
\begin{Bmatrix}J'&F'&I\\F&J&1\end{Bmatrix}
\begin{pmatrix}F&1&F'\\M&q&-M'\end{pmatrix}.
\end{eqnarray}
Here the array in the curly braces is a 6$j$ symbol, 
the array in the parentheses is a 3$j$ symbol, 
and $\langle J' \| D\| J\rangle$ is the reduced electric-dipole matrix element in the $J$ basis.
For the cesium $D_2$ line, we have $\langle J' \| D\| J\rangle=6.347$ a.u. 
$=5.38\times10^{-29}$ C m \cite{coolingbook}. 
We note that the spherical tensor components $d_{M'M}^{(q)}$
represent the dipole matrix-element vector $\mathbf{d}_{M'M}=\sum_q (-1)^q d_{M'M}^{(q)}\boldsymbol{\epsilon}_{-q}$ in the spherical basis 
$\{\boldsymbol{\epsilon}_{-1},\boldsymbol{\epsilon}_{0},\boldsymbol{\epsilon}_{1}\}$, where $\boldsymbol{\epsilon}_{-1}=(\hat{\mathbf{x}}-i\hat{\mathbf{y}})/\sqrt2$ and $\boldsymbol{\epsilon}_{1}=-(\hat{\mathbf{x}}+i\hat{\mathbf{y}})/\sqrt2$ are complex basis vectors and $\boldsymbol{\epsilon}_{0}=\hat{\mathbf{z}}$ is a real basis vector. 
It is clear that $\mathbf{d}_{M'M}$ is a real vector for $M'=M$ ($\pi$ transitions) and is a complex vector for $M'=M\pm1$ ($\sigma_{\pm}$ transitions).

We introduce the notation $\mathcal{E}_{q}$ with $q=0,\pm1$ for the spherical tensor components of the field envelope vector $\boldsymbol{\mathcal{E}}$,
that is, $\mathcal{E}_{-1}=(\mathcal{E}_x-i\mathcal{E}_y)/\sqrt{2}$, $\mathcal{E}_0=\mathcal{E}_z$, and $\mathcal{E}_{1}=-(\mathcal{E}_x+i\mathcal{E}_y)/\sqrt{2}$.
The interaction of the atom with the classical coherent probe field is characterized by 
the set of Rabi frequencies 
\begin{equation}\label{x4}
\Omega_{eg}=\frac{1}{\hbar}(\mathbf{d}_{eg}\cdot\boldsymbol{\mathcal{E}})
=\frac{1}{\hbar}\sum_{q=0,\pm1} (-1)^q d_{eg}^{(q)}\mathcal{E}_{-q}.
\end{equation}
The time evolution of the reduced density operator $\rho$ of the atom
is governed by the generalized Bloch equations \cite{absorption}
\begin{eqnarray}\label{x5}
\dot{\rho}_{ee'}&=&\frac{i}{2}\sum_{g}(\Omega_{eg}\rho_{ge'}
-\Omega_{e'g}^*\rho_{eg})\nonumber\\&&\mbox{}
-\frac{1}{2}\sum_{e''}(\gamma^{(\mathrm{tot})}_{ee''}{\rho}_{e''e'}+\gamma^{(\mathrm{tot})}_{e''e'}{\rho}_{ee''}),\nonumber\\
\dot{\rho}_{gg'}&=&-\frac{i}{2}\sum_{e}(\Omega_{eg'}
\rho_{ge}-\Omega_{eg}^*\rho_{eg'})
+\sum_{ee'}\gamma^{(\mathrm{tot})}_{e'eg'g}{\rho}_{ee'},\nonumber\\
\dot{\rho}_{eg}&=&i\delta_{eg}\rho_{eg}
+\frac{i}{2}\sum_{g'}\Omega_{eg'}\rho_{g'g}
-\frac{i}{2}\sum_{e'}\Omega_{e'g}\rho_{ee'}
\nonumber\\&&\mbox{} 
-\frac{1}{2}\sum_{e'} \gamma^{(\mathrm{tot})}_{ee'}\rho_{e'g}.
\end{eqnarray}
Here $\delta_{eg}=\omega-\omega_{eg}$ is the detuning of the field from the atomic transition frequency $\omega_{eg}=\omega_e-\omega_g$. In the case of the transitions between the Zeeman sublevels of the hfs levels $F'$ and $F$, we have $\omega_{eg}=\omega_0$ and $\delta_{eg}=\delta=\omega-\omega_0$. The coefficients $\gamma^{(\mathrm{tot})}_{ee'gg'}$ and $\gamma^{(\mathrm{tot})}_{ee'}$ characterize the effect of spontaneous emission on the reduced density operator of the atomic state. They are given as  \cite{cesium decay}
$\gamma^{(\mathrm{tot})}_{ee'gg'}=\gamma^{(\mathrm{gyd})}_{ee'gg'}+\gamma^{(\mathrm{rad})}_{ee'gg'}$
and $\gamma^{(\mathrm{tot})}_{ee'}=\sum_{g}\gamma^{(\mathrm{tot})}_{ee'gg}=\gamma^{(\mathrm{gyd})}_{ee'}+\gamma^{(\mathrm{rad})}_{ee'}$.
Here the set of coefficients $\gamma^{(\mathrm{gyd})}_{ee'gg'}$ and $\gamma^{(\mathrm{gyd})}_{ee'}=\sum_{g}\gamma^{(\mathrm{gyd})}_{ee'gg}$ describes spontaneous emission into guided modes, and the set of coefficients $\gamma^{(\mathrm{rad})}_{ee'gg'}$ and 
$\gamma^{(\mathrm{rad})}_{ee'}=\sum_{g}\gamma^{(\mathrm{rad})}_{ee'gg}$ describes spontaneous emission into radiation modes. 
The total decay rate of the population of the excited magnetic sublevel $|e\rangle$ is 
$\gamma^{(\mathrm{tot})}_{ee}=\gamma^{(\mathrm{gyd})}_{ee}+\gamma^{(\mathrm{rad})}_{ee}$.
The explicit expressions for the decay coefficients are given in Ref.~\cite{cesium decay} and are summarized in Appendices \ref{sec:guided} and  \ref{sec:radiation}.
 
We note that the density-matrix equations \eqref{x5} are consistent with those used in the treatments for the excitation of a multilevel atom by light of arbitrary polarization \cite{Milner98,Milner99,Taichenachev99,Vitanov03,Taichenachev04,Yudin13,ChangMinogin}. Equations \eqref{x5} can, in principle, be used for an arbitrary (degenerate and non-degenerate) multilevel atom.
The tensor nature of the Zeeman sublevels and the hfs levels of a realistic alkali-metal atom is expressed by Eq. \eqref{x3} for the spherical tensor components $d_{eg}^{(q)}$ of the atomic dipole matrix elements $\mathbf{d}_{eg}$. These quantities enter Eqs. \eqref{x5} through expression \eqref{x4} for the Rabi frequencies $\Omega_{eg}$.
Unlike the case of the atom--field system in free space \cite{ChangMinogin}, the presence of the nanofiber modifies the decay rates $\gamma^{(\mathrm{tot})}_{ee}$ and leads to the appearance of the cross-level decay coefficients $\gamma^{(\mathrm{tot})}_{e_1e_2}$ (with $e_1\not= e_2$) in Eqs.~\eqref{x5} (see \cite{cesium decay}).

\subsection{Scattering rates into the guided modes with given propagation directions and polarizations}
\label{subsec:scatt}

We assume that the atom is initially prepared in an incoherent mixture of the Zeeman sublevels 
$|M\rangle$ of the ground-state hyperfine level $F$ and that the initial population distribution of the atom is independent of $M$. 
We are interested in the regime where the probe field $\boldsymbol{\mathcal{E}}$ is stationary and the atom is in its steady state.

The total rate of scattering of incident photons from the atom is given by
$\Gamma_{\mathrm{tot}}=\sum_{ee'}\gamma^{(\mathrm{tot})}_{ee'}\rho_{e'e}=\Gamma_{\mathrm{gyd}}+\Gamma_{\mathrm{rad}}$,
where $\Gamma_{\mathrm{gyd}}=\sum_{ee'}\gamma_{ee'}^{\mathrm{(gyd)}}\rho_{e'e}$
and $\Gamma_{\mathrm{rad}}=\sum_{ee'}\gamma_{ee'}^{\mathrm{(rad)}}\rho_{e'e}$
are the rates of scattering into guided modes and radiation modes, respectively. 

We are interested in the rate of scattering into the guided modes with a given propagation direction and a given polarization. 
We consider not only the quasicircular polarizations with the index $l=+,-$ but also the quasilinear polarizations with the index $\xi=x,y$. To combine the two cases, we use the notation $p$ that can be either $l=+,-$ or $\xi=x,y$. The description of the structure of the guided field is given in \cite{fibermode,fiber books} and is summarized in Appendices \ref{sec:guided} and  \ref{sec:radiation}.
The set of modes with quasicircular polarizations $l=+,-$ and the set of modes with quasilinear polarizations $\xi=x,y$ can be used each to define an orthogonal mode basis for the fundamental guided modes HE$_{11}$ \cite{fiber books}. These two sets overlap each other and have different properties. They are chosen not only for convenience but also to describe the experimental situations where a guided probe field with a given polarization is employed and/or
a scattered field with a given polarization is measured \cite{Reitz14}.

The rate of scattering into the guided modes with a given propagation direction $f=\pm$ and a given polarization $p=+$, $-$, $x$, or $y$ is given by
\begin{equation}\label{x6a} 
\Gamma_{fp}=\sum_{ee'}\gamma_{ee'}^{(fp)}\rho_{e'e}.
\end{equation}
Here we have introduced the notation 
\begin{equation}\label{x6b} 
\gamma_{ee'}^{(fp)}=\gamma_{ee'}^{(fp fp)}, 
\end{equation}
where
\begin{equation}\label{x6}
\gamma_{ee'}^{(fp f'p')}=2\pi \sum_{g}G_{\omega_0fp eg}G_{\omega_0f'p' e'g}^*.
\end{equation}
In the above expression, $G_{\omega_0fp eg}$ is the coefficient for the coupling between the resonant guided mode $\omega_0fp$ and the atomic transition $|e\rangle\leftrightarrow |g\rangle$.
For quasicircularly polarized modes, that is, for $p=l=\pm$, the general expression for the coefficient $G_{\omega fp eg}=G_{\omega fl eg}$ is given by Eq.~(\ref{g13}) in Appendix \ref{sec:guided}, which reads
\begin{eqnarray}\label{x7}
G_{\omega fl eg}=\sqrt{\frac{\omega\beta'}{4\pi\epsilon_0\hbar}}\;
\big(\mathbf{d}_{eg}\cdot\mathbf{e}^{(\omega fl)}\big)e^{i(f\beta z+l\varphi)}.
\end{eqnarray}
Here $\mathbf{e}^{(\omega fl)}$ is the normalized profile function for quasicircularly polarized guided modes and is given by Eqs.~(\ref{g6}) and  (\ref{g7}). 
For quasilinearly polarized modes, that is, for $p=\xi=x,y$, the general expression for the coefficient $G_{\omega fp eg}=G_{\omega f\xi eg}$ is given by Eq.~(\ref{g14}) in Appendix \ref{sec:guided}, which reads 
\begin{eqnarray}\label{x8}
G_{\omega f\xi eg}=\sqrt{\frac{\omega\beta'}{4\pi\epsilon_0\hbar}}\;
\big(\mathbf{d}_{eg}\cdot\mathbf{e}^{(\omega f\xi)}\big)e^{if\beta z}.
\end{eqnarray}
Here $\mathbf{e}^{(\omega f\xi)}$ is the normalized profile function for quasilinearly polarized guided modes and is given by Eqs.~(\ref{g9}). 

We introduce the notation
\begin{equation}\label{x8a}
\gamma_{ee'}^{(f)}=\sum_{l=+,-}\gamma_{ee'}^{(fl)}=\sum_{\xi=x,y}\gamma_{ee'}^{(f\xi)}
\end{equation}
for the coefficients of spontaneous emission into guided modes in a given direction $f$.
We use the abbreviations $\gamma_{ee'}^{(\mathrm{fw})}=\gamma_{ee'}^{(+)}$ and $\gamma_{ee'}^{(\mathrm{bw})}=\gamma_{ee'}^{(-)}$.
We have the relation $\gamma_{ee'}^{(\mathrm{gyd})}=\gamma_{ee'}^{(\mathrm{fw})}+\gamma_{ee'}^{(\mathrm{bw})}$.

Similarly, we introduce the notation
\begin{equation}\label{x8b}
\Gamma_{f}=\sum_{l=+,-}\Gamma_{fl}=\sum_{\xi=x,y}\Gamma_{f\xi}
\end{equation}
for the rate of scattering into guided modes in a given direction $f$.
We use the abbreviations $\Gamma_{\mathrm{fw}}=\Gamma_{+}$ and $\Gamma_{\mathrm{bw}}=\Gamma_{-}$.
The scattering rate into guided modes $\Gamma_{\mathrm{gyd}}$ can be decomposed as 
$\Gamma_{\mathrm{gyd}}=\Gamma_{\mathrm{fw}}+\Gamma_{\mathrm{bw}}$.

We note that the output guided probe field is the result of the interference between the input guided probe field and the field scattered into the forward guided modes.
It follows from the energy conservation law that the loss rate for the input guided probe field is given by 
\begin{equation}\label{x10c}
\Gamma_{\mathrm{loss}}=\Gamma_{\mathrm{rad}}+\Gamma_{\mathrm{bw}}.
\end{equation}

The total rate of scattering $\Gamma_{\mathrm{tot}}$ and its components $\Gamma_{\mathrm{gyd}}$ and $\Gamma_{\mathrm{rad}}$ have been calculated systematically in Ref.~\cite{absorption}. In this earlier work, the rates $\Gamma_{\mathrm{fw}}$, $\Gamma_{\mathrm{bw}}$, and $\Gamma_{\mathrm{loss}}$ have been deduced from $\Gamma_{\mathrm{gyd}}$ and $\Gamma_{\mathrm{rad}}$ by using the incorrect formula $\Gamma_{\mathrm{fw}}=\Gamma_{\mathrm{bw}}$. The latter relation is valid in the cases of the far field emitted from a two-level atom with a real dipole matrix-element vector in free space \cite{Loudon,Scully,Mandel} or in the presence of a waveguide \cite{Domokos02}. For a two-level atom with a complex dipole matrix-element vector or 
a multilevel atom in the vicinity of a fiber, the formula  $\Gamma_{\mathrm{fw}}=\Gamma_{\mathrm{bw}}$ is, as shown below, not valid. 

Indeed, it follows from Eq.~(\ref{g23}) that, when the fiber axis $z$ is used as the quantization axis for the internal state of an alkali-metal atom, the coefficients $\gamma_{ee'}^{(fp)}$ and $\gamma_{ee'}^{(\bar{f}p)}$  of spontaneous decay into the guided modes with the opposite propagation directions $f$ and $\bar{f}=-f$, respectively, satisfy the relation
$\gamma_{ee'}^{(fp)}=(-1)^{M_e-M_{e'}}\gamma_{ee'}^{(\bar{f}p)}$.
When we take $e'=e$, $e\pm1$, or $e\pm2$, we get the symmetry relations
\begin{subequations}\label{x11}
\begin{eqnarray} 
\gamma_{ee}^{(fp)}&=&\gamma_{ee}^{(\bar{f}p)},\label{x11a}\\ 
\gamma_{e,e\pm1}^{(fp)}&=&-\gamma_{e,e\pm1}^{(\bar{f}p)},\label{x11b}\\ 
\gamma_{e,e\pm2}^{(fp)}&=&\gamma_{e,e\pm2}^{(\bar{f}p)},\label{x11c}
\end{eqnarray} 
\end{subequations}
respectively. All other decay coefficients, i.e., the coefficients $\gamma_{ee'}^{(fp)}$ with $e'\not=e,e\pm1,e\pm2$, are equal to zero  due to the transition selection rules. When the atom has a single upper level $|e\rangle$, only the diagonal coefficient
$\gamma_{ee}^{(fp)}$ appears in the expression for the scattering rate $\Gamma_{fp}$. In this case, due to the relation $\gamma_{ee}^{(fp)}=\gamma_{ee}^{(\bar{f}p)}$ [see Eq.~(\ref{x11a})], we have the equalities $\Gamma_{fp}=\Gamma_{\bar{f}p}$ and, consequently, $\Gamma_{\mathrm{fw}}=\Gamma_{\mathrm{bw}}$ for the scattering rates in the opposite directions. However, when the atom has two or more upper levels, the expression for the scattering rate $\Gamma_{fp}$ contains not only the diagonal coefficients
$\gamma_{ee}^{(fp)}$ but also the off-diagonal coefficients $\gamma_{e,e\pm1}^{(fp)}$. 
Note that $\gamma_{e,e\pm1}^{(fp)}=2\pi \sum_{g}G_{\omega_0fp,e,g}G_{\omega_0fp,e\pm1,g}^*$. 
Since the guided-mode profile function $\mathbf{e}^{(\omega fl)}$ has
a longitudinal component $e_z^{(\omega fl)}$ in addition to the transverse components $e_{r}^{(\omega fl)}$ and 
$e_{\varphi}^{(\omega fl)}$, we may have $\gamma_{e,e\pm1}^{(fp)}\not=0$. This fact and   
the relation $\gamma_{e,e\pm1}^{(fp)}=-\gamma_{e,e\pm1}^{(\bar{f}p)}$  [see Eq.~(\ref{x11b})] may lead to $\Gamma_{fp}\not=\Gamma_{\bar{f}p}$ and, hence, to $\Gamma_{\mathrm{fw}}\not=\Gamma_{\mathrm{bw}}$. Thus, the rates of scattering of light from a multilevel atom into the guided modes of a nanofiber in the forward and backward directions may differ from each other. 
Such asymmetry is a result of the complexity of the atomic level and transition structures and the existence of a longitudinal component of the guided-mode profile function. 
We note that Eqs.~\eqref{x11} are valid only in the case where the fiber axis $z$ is used as the quantization axis for the atomic internal states
and the atomic transitions $|e\rangle\leftrightarrow |g\rangle$ are of the type $\pi$, $\sigma_+$, or $\sigma_-$ with respect to this specific quantization axis. 
We emphasize that, when we perform the summation over the Zeeman sublevels of the atom with the initial $M$-independent distribution of populations, we obtain the answers that are independent of the choice of the quantization axis. 

For a two-level atom with a real dipole matrix-element vector
or a complex dipole matrix element of the  $\sigma_{\pm}$ transition type with respect to the fiber axis $z$,
we have $\gamma_{ee}^{(\mathrm{fw})}=\gamma_{ee}^{(\mathrm{bw})}$ and, hence, $\Gamma_{\mathrm{fw}}=\Gamma_{\mathrm{bw}}$.
For a two-level atom with an arbitrary dipole matrix-element vector $\mathbf{d}$,  the relations $\gamma_{ee}^{(\mathrm{fw})}=\gamma_{ee}^{(\mathrm{bw})}$ and $\Gamma_{\mathrm{fw}}=\Gamma_{\mathrm{bw}}$ are, in general, not valid. We will show in Sec. \ref{sec:twolevelatom} that
we may obtain $\gamma_{ee}^{(\mathrm{fw})}\not=\gamma_{ee}^{(\mathrm{bw})}$ and, consequently, 
$\Gamma_{\mathrm{fw}}\not=\Gamma_{\mathrm{bw}}$ for a two-level atom with a transition of the $\sigma_{\pm}$ type with respect to the $y$ axis.

The differences between the rates of the forward and backward scattering processes are
\begin{eqnarray}\label{x12}
\Gamma_{fp}-\Gamma_{\bar{f}p}&=&4\mathrm{Re}\sum_{e}\gamma_{e,e+1}^{(fp)}\rho_{e+1,e},\nonumber\\
\Gamma_{\mathrm{fw}}-\Gamma_{\mathrm{bw}}&=& \pm 4\mathrm{Re}\sum_{e}\gamma_{e,e+1}^{(\pm)}\rho_{e+1,e}.
\end{eqnarray}
The above expressions show clearly that the difference between the forward and backward scattering  processes is caused
by the interference between the $\pi$ and $\sigma_\pm$ downward transitions of the atom. 
Such interference may be constructive or destructive depending on the scattering direction.
The interference between the downward transitions from the levels $M_e$ and $M_e\pm1$ 
may appear only if the spontaneous emission coefficients $\gamma_{e,e\pm1}^{(fp)}$ as well as the off-diagonal density matrix elements $\rho_{e\pm1,e}$ are not zero.

We note that, for an arbitrary position on the $x$ axis, the longitudinal component $e_z^{(\omega fy)}$ of the profile function $\mathbf{e}^{(\omega fy)}$ of the $y$-polarized guided modes 
is vanishing. Consequently, when the atom is positioned on the $x$ axis, we have $\gamma_{e,e\pm1}^{(fy)}=0$. This leads to 
\begin{equation}\label{x13}
\Gamma_{fy}=\Gamma_{\bar{f}y}.
\end{equation}
Thus, the rate of scattering of light from the atom into the quasilinearly $y$-polarized guided modes, where the principal polarization direction $y$ is perpendicular to the radial direction $x$ of the atomic position,
does not depend on the scattering direction $f$. Due to this property, 
the difference between the rates of scattering into the forward and backward guided modes is 
\begin{equation}\label{x13c}
\Delta\Gamma_{\mathrm{fwbw}}\equiv\Gamma_{\mathrm{fw}}-\Gamma_{\mathrm{bw}}=\Gamma_{+,x}-\Gamma_{-,x}.
\end{equation}

To make the rates of scattering into forward and backward guided modes different from each other, not only the spontaneous emission coefficients $\gamma_{e,e\pm1}^{(fp)}$ but also the off-diagonal density matrix elements $\rho_{e,e\pm1}$ must not be all equal to zero. Nonzero coherence between neighboring Zeeman levels $M_e$ and $M_e\pm1$ of the atom can be induced only 
if the probe field $\boldsymbol{\mathcal{E}}$, which is general (not necessarily a guided light field) in this discussion, has a nonzero component $\mathcal{E}_z$ along the fiber axis $z$, that is,
\begin{equation}
\mathcal{E}_z\not=0.
\end{equation}
In addition, the presence of a nonzero component $\mathcal{E}_\perp$ of the probe field $\boldsymbol{\mathcal{E}}$ in the fiber transverse plane $xy$
is also required. 
Moreover, the phases of the off-diagonal density matrix elements $\rho_{e,e\pm1}$ must be appropriate so that the interference described by the terms
$\mathrm{Re}(\gamma_{e,e+1}^{(\cdots)}\rho_{e+1,e})$ is not washed out by the summation over $e$ in Eqs.~(\ref{x12}). 

We can show that, when the probe field $\boldsymbol{\mathcal{E}}$ has no component along the radial direction $x$ of the atomic position, that is, when 
\begin{equation}\label{x13a}
\mathcal{E}_x=0, 
\end{equation}
we have the properties $\mathcal{E}_q=\mathcal{E}_{-q}$ (with $q=0,\pm1$) and, consequently, $\Omega_{eg}=(-1)^{F-F'+1}\Omega_{\bar{e}\bar{g}}$. 
Here we have introduced the notations $\bar{e}$ and $\bar{g}$
for the Zeeman sublevels with the magnetic quantum numbers $-M_e$ and $-M_g$, respectively, of the excited-state hyperfine level $F'$ and the ground-state hyperfine level $F$,
respectively. Then, we obtain the relation $\rho_{ee'}=\rho_{\bar{e}\bar{e'}}$ in the steady-state regime. 
In deriving this relation we have used Eqs.~\eqref{x5} and the symmetry properties
of the spontaneous emission coefficients given in Appendices \ref{sec:guided} and \ref{sec:radiation}.
On the other hand, since the atom is on the axis $x$, Eq.~(\ref{g36}) yields $\gamma_{ee'}^{(flfl')}=(-1)^{M_e-M_{e'}}\gamma_{\bar{e}\bar{e'}}^{(f\bar{l}f\bar{l'})}$, where $l=\pm$ and $\bar{l}=-l=\mp$. Hence we find that the interference terms $\mathrm{Re}(\gamma_{e,e+1}^{(fx)}\rho_{e+1,e})$ for the scattering into the guided modes with the quasilinear polarization $x$ are washed out by the summation over $e$ in the steady-state regime.
Consequently, we have $\Gamma_{fx}=\Gamma_{\bar{f}x}$. This equality and the equality $\Gamma_{fy}=\Gamma_{\bar{f}y}$, which is valid for arbitrary polarization of the probe field $\boldsymbol{\mathcal{E}}$, give $\Gamma_{\mathrm{fw}}=\Gamma_{\mathrm{bw}}$. Thus, under the condition \eqref{x13a} and in the steady-state regime, 
the forward and backward scattering processes have the same rates.
In addition, we obtain the equality $\Gamma_{fl}=\Gamma_{\bar{f} \bar{l}}$ 
for $l=\pm$ for the rates of scattering into the guided modes with the opposite
circular polarizations $l=\pm$ and $\bar{l}=-l$ in the opposite directions $f=\pm$ and $\bar{f}=-f$. 
When we combine the result of the above discussion with the result of the discussion in the previous paragraph, we see that 
both components $\mathcal{E}_z$ and $\mathcal{E}_x$ of the probe field must be nonzero to produce a difference between the forward and backward scattering rates. 

Our additional analysis shows that, when the probe field $\boldsymbol{\mathcal{E}}$ has no component along the direction $y$, which is perpendicular to the radial direction $x$ of the atomic position, that is, when 
\begin{equation}\label{x13b}
\mathcal{E}_y=0, 
\end{equation}
we have the properties $\mathcal{E}_q=(-1)^q\mathcal{E}_{-q}$ (with $q=0,\pm1$) and, hence, $\Omega_{eg}=(-1)^{F-F'+1+M_e-M_g}\Omega_{\bar{e}\bar{g}}$. 
Using these properties and the symmetry properties of the spontaneous emission coefficients, we find 
from Eqs.~\eqref{x5} the relation $\rho_{ee'}=(-1)^{M_e-M_{e'}}\rho_{\bar{e}\bar{e'}}$ for the populations of and coherences between the Zeeman sublevels of the excited state in the steady-state regime. Then, the difference $\Delta\Gamma_{\mathrm{fwbw}}$ between the rates of scattering into the forward and backward guided modes is, in general, nonvanishing, unlike the result in the case \eqref{x13a}. In the particular case where the probe field 
$\boldsymbol{\mathcal{E}}$ is elliptically polarized in the $zx$ plane, the rate difference $\Delta\Gamma_{\mathrm{fwbw}}$ can become significantly different from zero. When the probe field $\boldsymbol{\mathcal{E}}$ is exactly linearly polarized in the $zx$ plane, the difference $\Delta\Gamma_{\mathrm{fwbw}}$ is zero for $\delta=0$ but may be slightly different from zero for $\delta\not=0$, depending
on the orientation of the field polarization vector $\mathbf{u}$ in the $zx$ plane.

\subsection{Numerical results}
\label{subsec:numer}

We solve the density-matrix equations (\ref{x5}) in the steady-state regime and use the results to
calculate the efficiency coefficients $\eta_{\mathrm{fw}}=\hbar\omega\Gamma_{\mathrm{fw}}/P_z$ 
and $\eta_{\mathrm{bw}}=\hbar\omega\Gamma_{\mathrm{bw}}/P_z$ for scattering into guided modes in the forward and backward directions, respectively. 
Here $P_z$ is the propagation power.
We also calculate the efficiency coefficients $\eta_{fp}=\hbar\omega\Gamma_{fp}/P_z$ for scattering into the guided modes with a given propagation direction $f$ and a given polarization $p$. 
In the numerical calculations, we use the fiber radius $a=250$ nm. 
As already stated, we consider the transitions between the hyperfine levels $6S_{1/2}F=4$ and $6P_{3/2}F'=5$ of the $D_2$ line of atomic cesium, with the wavelength $\lambda_0=852$ nm. 
The atom is positioned on the positive side of the axis $x$. The propagation power of the guided probe light field is assumed to be $P_z=10$ fW. This power is much lower than the saturation power $P_{\mathrm{sat}}=4.4$ pW. Here $P_{\mathrm{sat}}$ is estimated as the power of a quasicircularly polarized guided light field that produces the intensity 
$I\equiv c\epsilon_0|\mathcal{E}|^2/2=I_{\mathrm{sat}}$ on the fiber surface, where $I_{\mathrm{sat}}=1.1$ mW/cm$^2$ is
the saturation intensity for a cesium atom with the cycling transition \cite{coolingbook}.
The power of 10 fW of the probe field corresponds to a flux of one photon per $\tau_{\mathrm{photon}}\simeq 23$ $\mu$s.
The corresponding value of the Rabi period for the cycling transition of a cesium atom on the fiber surface
is $\tau_{\mathrm{Rabi}}\simeq 6$ $\mu$s. The characteristic times $\tau_{\mathrm{photon}}$ and $\tau_{\mathrm{Rabi}}$ are short as compared to the experimentally observed trapping lifetime $\tau_{\mathrm{trap}}\simeq 100$ ms for atoms in the two-color fiber-based trap \cite{Vetsch10}. 
Note that the power of 10 fW of the guided probe field is low enough that
the hyperfine pumping is negligible in the interaction process \cite{Siddons08}. Indeed, a simple estimate shows that the off-resonant hyperfine scattering rate for a cesium atom on the fiber surface is on the order of $4$ s$^{-1}$. It is clear that the effect of the off-resonant hyperfine scattering is very small in the characteristic times $\tau_{\mathrm{photon}}$ and $\tau_{\mathrm{Rabi}}$ and
is small in the trapping lifetime $\tau_{\mathrm{trap}}$. 

\begin{figure}[tbh]
\begin{center}
  \includegraphics{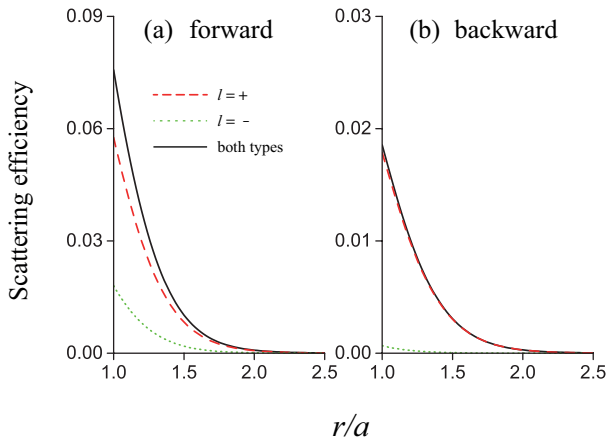}
 \end{center}
\caption{(Color online) Radial-distance dependencies of the scattering efficiency coefficients in the case where the guided probe field $\boldsymbol{\mathcal{E}}$ is
quasicircularly polarized. 
The principal circulation direction of the polarization of the probe field is counterclockwise.
The coefficients $\eta_{\mathrm{fw}}$ 
and $\eta_{\mathrm{bw}}$ for scattering into forward and backward guided modes, respectively, are shown by the solid black lines
in parts (a) and (b), respectively. 
The coefficients $\eta_{fl}$ with $l=+$ (dashed red lines) and $l=-$ (dotted green lines)
for the modes with the individual counterclockwise and clockwise polarizations, respectively, are also shown. 
The fiber radius is $a=250$ nm, the light wavelength is $\lambda=852$ nm, 
and the light propagation power is $P_z=10$ fW. The detuning of the field frequency from the atomic transition frequency is $\delta=0$.
}
\label{fig2}
\end{figure}

\begin{figure}[tbh]
\begin{center}
  \includegraphics{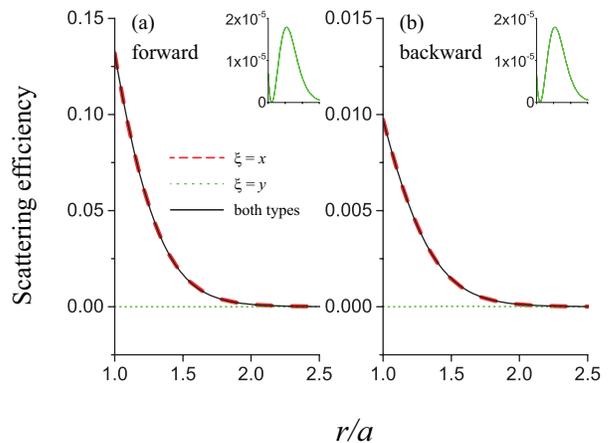}
 \end{center}
\caption{(Color online) 
Radial-distance dependencies of the scattering efficiency coefficients in the case where the guided probe field $\boldsymbol{\mathcal{E}}$ is
quasilinearly polarized along the direction of the axis $x$, on which the atom is located. 
The coefficients $\eta_{\mathrm{fw}}$ 
and $\eta_{\mathrm{bw}}$ for scattering into forward and backward guided modes, respectively, are shown by the solid black lines
in parts (a) and (b), respectively. 
The coefficients $\eta_{f\xi}$ with $\xi=x$ (dashed red lines) and $\xi=y$ (dotted green lines)
for the modes with the individual $x$ and $y$ polarizations, respectively, are also shown.
The insets in parts (a) and (b) show the details of $\eta_{f\xi}$ with $\xi=y$. 
Other parameters are as in Fig.~\ref{fig2}.
}
\label{fig3}
\end{figure}

\begin{figure}[tbh]
\begin{center}
  \includegraphics{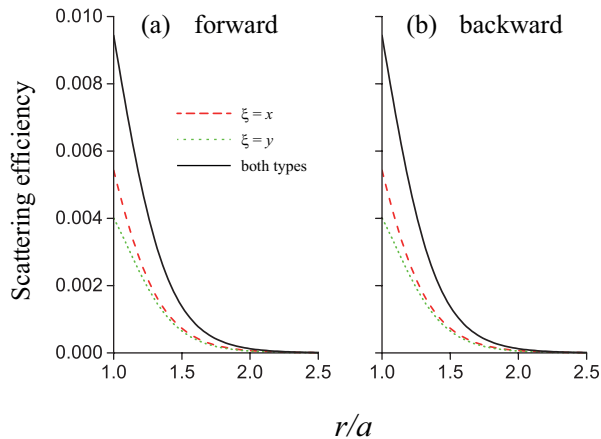}
 \end{center}
\caption{(Color online) 
Same as Fig.~\ref{fig3} but the guided probe field $\boldsymbol{\mathcal{E}}$ is
quasilinearly polarized along the axis $y$, which is perpendicular to the radial direction of the position of the atom. 
}
\label{fig4}
\end{figure}

We plot in Figs.~\ref{fig2}, \ref{fig3}, and \ref{fig4} the dependencies of the scattering efficiency coefficients on the normalized radial distance $r/a$
in the cases where the guided probe field $\boldsymbol{\mathcal{E}}$ is quasicircularly polarized, $x$-polarized, and $y$-polarized, respectively.
A common feature of these figures is that, in general (except for $\eta_{fy}$ in the case of Fig.~\ref{fig3}),
the scattering efficiency coefficients reduce with increasing radial distance $r$.
Such a reduction is due to the evanescent-wave profiles of the guided-mode functions outside the fiber, which affect
the scattering rates $\Gamma_{fp}$ via the Rabi frequencies $\Omega_{eg}$ and the spontaneous emission coefficients $\gamma_{ee'}^{(fp)}$.

\begin{figure}[tbh]
\begin{center}
  \includegraphics{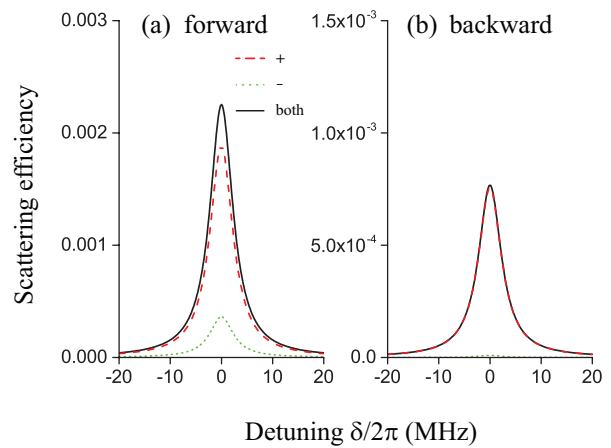}
 \end{center}
\caption{(Color online) Frequency dependencies of the scattering efficiency coefficients in the case where the guided probe field $\boldsymbol{\mathcal{E}}$ is
quasicircularly polarized. 
The principal circulation direction of the polarization of the probe field is counterclockwise.
The coefficients $\eta_{\mathrm{fw}}$ 
and $\eta_{\mathrm{bw}}$ for scattering into forward and backward guided modes, respectively, are shown by the solid black lines
in parts (a) and (b), respectively. 
The coefficients $\eta_{fl}$ with $l=+$ (dashed red lines) and $l=-$ (dotted green lines)
for the modes with the individual counterclockwise and clockwise polarizations, respectively, are also shown. 
The radial position of the atom is $r/a=1.8$.
Other parameters are as in Fig.~\ref{fig2}.
}
\label{fig5}
\end{figure}

\begin{figure}[tbh]
\begin{center}
  \includegraphics{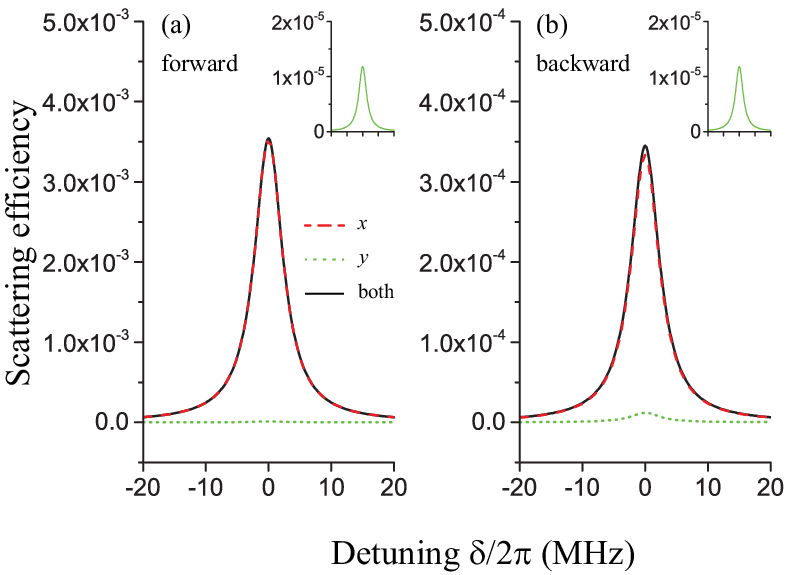}
 \end{center}
\caption{(Color online) 
Frequency dependencies of the scattering efficiency coefficients in the case where the guided probe field $\boldsymbol{\mathcal{E}}$ is
quasilinearly polarized along the direction of the axis $x$, on which the atom is located. 
The coefficients $\eta_{\mathrm{fw}}$ 
and $\eta_{\mathrm{bw}}$ for scattering into forward and backward guided modes, respectively, are shown by the solid black lines
in parts (a) and (b), respectively. 
The coefficients $\eta_{f\xi}$ with $\xi=x$ (dashed red lines) and $\xi=y$ (dotted green lines)
for the modes with the individual $x$ and $y$ polarizations, respectively, are also shown.
The insets in parts (a) and (b) show the details of $\eta_{f\xi}$ with $\xi=y$. 
The radial position of the atom is $r/a=1.8$.
Other parameters are as in Fig.~\ref{fig2}.
}
\label{fig6}
\end{figure}

\begin{figure}[tbh]
\begin{center}
  \includegraphics{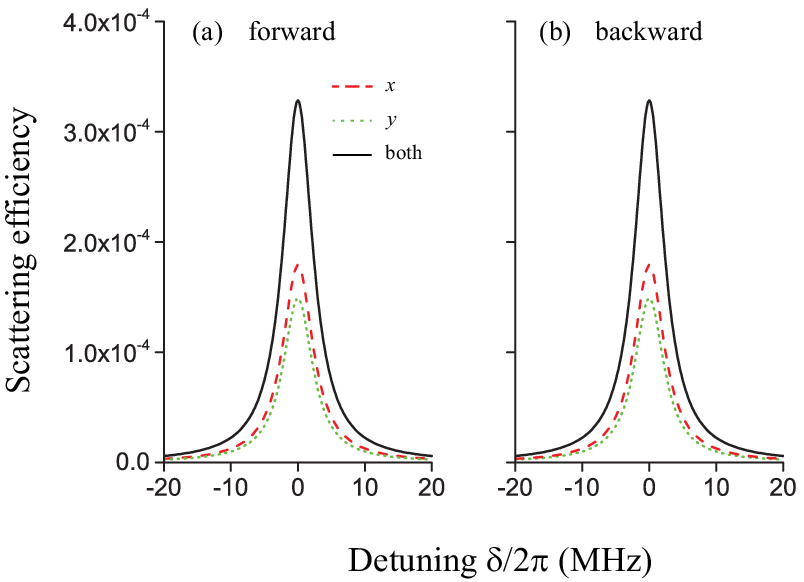}
 \end{center}
\caption{(Color online) 
Same as Fig.~\ref{fig6} but the guided probe field $\boldsymbol{\mathcal{E}}$ is
quasilinearly polarized along the axis $y$, which is perpendicular to the radial direction of the position of the atom. 
}
\label{fig7}
\end{figure}

We plot in Figs.~\ref{fig5}, \ref{fig6}, and \ref{fig7} the dependencies of the scattering efficiency coefficients on the field detuning  $\delta$  
in the cases where the guided probe field $\boldsymbol{\mathcal{E}}$ is quasicircularly polarized, $x$-polarized, and $y$-polarized, respectively.
A common feature of the plotted curves is that they look like Lorentzian lines, with a peak at the resonance frequency $\omega=\omega_0$. We note that the linewidth of the calculated curves is about $5.4$ MHz. This value is slightly larger than the literature value of $5.2$ MHz for the atomic natural (free-space) linewidth \cite{coolingbook} although the power of the field is very low and hence the effect of power broadening is very weak. The numerically observed broadening depends on the radial position of the atom and is due to the enhancement of spontaneous emission by the fiber \cite{cesium decay}. For the parameters used, the enhancement factor is about $1.03$.

Figures \ref{fig2} and \ref{fig5} show the radial-distance and field-frequency dependencies of the scattering efficiency coefficients in the case where the guided probe field $\boldsymbol{\mathcal{E}}$ is counterclockwise quasicircularly polarized. We note that the case where the field is clockwise polarized is similar to the case where the field is counterclockwise polarized. 
Comparison between parts (a) and (b) of Fig.~\ref{fig2} and between that of Fig.~\ref{fig5} 
shows that, when the probe field is quasicircularly polarized, the efficiency of the scattering into the forward guided modes is a few times larger than that of the scattering into the backward guided modes. We observe from Figs.~\ref{fig2} and \ref{fig5} the appearance of a new polarization component, namely the clockwise polarization (see the dotted green lines), of the field in the guided modes. The relative magnitude of the secondary polarization field component is significant in the forward modes but not substantial in the backward modes.

Figures \ref{fig3} and \ref{fig6} show the radial-distance and field-frequency dependencies of the scattering efficiency coefficients in the case where the guided probe field $\boldsymbol{\mathcal{E}}$ is 
quasilinearly polarized along the axis $x$.
Comparison between parts (a) and (b) of Fig.~\ref{fig3} and between that of Fig.~\ref{fig6} shows that, when the probe field is 
quasilinearly polarized along the axis $x$, the efficiency of the scattering into the forward guided modes is about one order of magnitude
stronger than that of the scattering into the backward guided modes.
We observe from Figs.~\ref{fig3} and \ref{fig6} that the main component of the total light scattered into the guided modes in the forward or backward direction (solid black lines) 
is the component with the polarization $x$ (dashed red lines), which is the same as the polarization of the probe field.
The insets of the figures show that the magnitude of the scattering efficiency of the $y$-polarized component (dotted green lines) is independent of the scattering direction and is three (in the case of backward direction) or four (in the case of forward direction) orders smaller than that of the $x$-polarized component (dashed red lines). 
The observed independence of the parameter $\eta_{fy}$ from the scattering direction $f$ is in agreement with Eq.~(\ref{x13}). 
It is a consequence of the fact that the axial component of the mode profile function of
the $y$-polarized guided modes is zero for the positions on the $x$ axis.
We observe from the insets of Fig.~\ref{fig3} that 
the scattering efficiency coefficient $\eta_{fy}$ increases with increasing $r/a$ in the interval from $\sim1.1$ to $\sim1.5$. 
Such an increase is different from the typical behavior of  the scattering efficiency coefficients and is
a result of interference between different channels of atomic transitions.

Figures \ref{fig4} and \ref{fig7} show the radial-distance and field-frequency dependencies of the scattering efficiency coefficients in the case where the guided probe field $\boldsymbol{\mathcal{E}}$ is 
quasilinearly polarized along the axis $y$. Comparison between parts (a) and (b) of Fig.~\ref{fig4} and between that of Fig.~\ref{fig7} shows that these parts are identical to each other. This means that, 
when the probe field is quasilinearly polarized along the axis $y$, the rates $\Gamma_{fx}$ and $\Gamma_{fy}$ of scattering into the guided modes with the individual $x$ and $y$ polarizations, respectively, and the total rate $\Gamma_f=\Gamma_{fx}+\Gamma_{fy}$ for both types of polarizations in a given direction $f$ are independent of the scattering direction $f=\pm$. This behavior is a consequence of the fact that the axial component $\mathcal{E}_z$ of the probe field $\boldsymbol{\mathcal{E}}$ is zero in the case considered. We observe from Figs.~\ref{fig4} and \ref{fig7} that both components with the $x$ (dashed red lines) and $y$ (dotted green lines) polarizations are present in 
the forward- and backward-scattered light fields (solid black lines). Furthermore, we note that the magnitude of the $x$-polarized component (dashed red lines) of the scattered field is slightly larger than that of 
the $y$-polarized component (dotted green lines) although the $y$ polarization is the polarization of the incident probe field. 
This feature is a consequence of the differences between  
the $x$- and $y$-polarized guided modes at the position of the atom, the properties of the atomic dipole matrix elements, and the properties of the atomic steady-state density matrix.

\begin{figure}[tbh]
\begin{center}
  \includegraphics{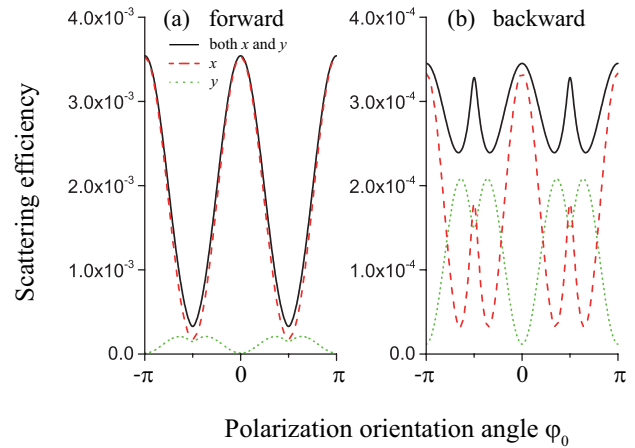}
 \end{center}
\caption{(Color online) 
Dependencies of the scattering efficiency coefficients on the azimuthal angle $\varphi_0$ of the polarization vector $\mathbf{u}$ of the guided probe field in the case where the latter is quasilinearly polarized. The coefficients $\eta_{\mathrm{fw}}$ and $\eta_{\mathrm{bw}}$ for scattering into forward and backward guided modes, respectively, are shown by the solid black lines in parts (a) and (b), respectively. 
The coefficients $\eta_{f\xi}$ with $\xi=x$ (dashed red lines) and $\xi=y$ (dotted green lines)
for the modes with the individual $x$ and $y$ polarizations, respectively, are also shown.
The atom is located on the $x$ axis at the distance $r/a=1.8$.
Other parameters are as in Fig.~\ref{fig2}.
}
\label{fig8}
\end{figure}
 
We plot in Fig.~\ref{fig8} the dependencies of the scattering efficiency coefficients on the azimuthal  angle $\varphi_0$ of the polarization vector $\mathbf{u}$ of the guided probe field in the case where the latter is quasilinearly polarized.
The figure shows that the scattering efficiency coefficients $\eta_{f\xi}$ are symmetric functions of $\varphi_0$, i.e., $\eta_{f\xi}(\varphi_0)=\eta_{f\xi}(-\varphi_0)$. In addition, we have $\eta_{f\xi}(\varphi_0)=\eta_{f\xi}(\pi\pm\varphi_0)$. 
These properties are obvious consequences of the symmetry of the atom--fiber system with respect to the radial direction of the atomic position in the fiber transverse plane.
We observe from the figure that the scattering efficiency coefficients vary significantly when we vary the relative orientation of the field polarization vector $\mathbf{u}$ with respect to the radial direction of the atomic position. Figure  \ref{fig8}(a) shows that, in the forward direction, the $x$-polarized component (dashed red line) of the scattered field is always larger than the $y$-polarized component (dotted green line). However, in the backward direction, according to Fig.~\ref{fig8}(b),
the $y$-polarized component (dotted green line) becomes larger than  the $x$-polarized component (dashed red line)
in two intervals of $\varphi_0$ in the region $-\pi/2\leq\varphi_0\leq\pi/2$ (in four intervals  of $\varphi_0$ 
in the region $-\pi\leq\varphi_0\leq\pi$). When we take into account the difference between the scales of the vertical axes and closely inspect the dotted green curves in parts (a) and (b) of Fig.~\ref{fig8},
we see that the scattering efficiency coefficient $\eta_{fy}$ does not depend on the scattering direction $f$, that is, we have $\eta_{fy}=\eta_{\bar{f}y}$, in agreement with Eq.~(\ref{x13}).

\begin{figure}[tbh]
\begin{center}
  \includegraphics{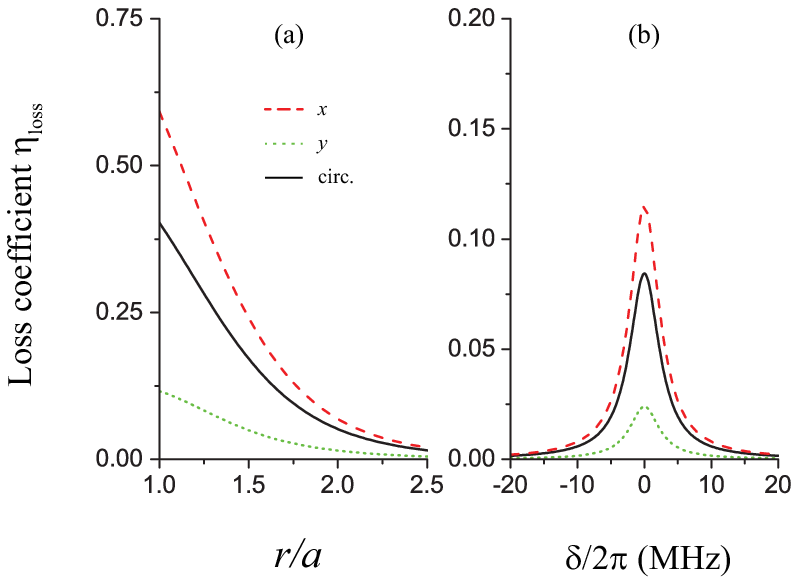}
 \end{center}
\caption{(Color online) 
Dependencies of the loss coefficient $\eta_{\mathrm{loss}}$ on the radial position of the atom (a) and the detuning of the guided probe field (b)
in the cases where the probe field is quasicircularly polarized (solid black lines),
quasilinearly polarized along the $x$ direction (dashed red lines), and quasilinearly polarized along the $y$ direction (dotted green lines). 
The atom is located on the $x$ axis. 
In (a), the field is tuned to exact resonance with the atom.
In (b), the radial position of the atom is $r/a=1.8$.
Other parameters are as in Fig.~\ref{fig2}.
}
\label{fig9}
\end{figure}

\begin{figure}[tbh]
\begin{center}
  \includegraphics{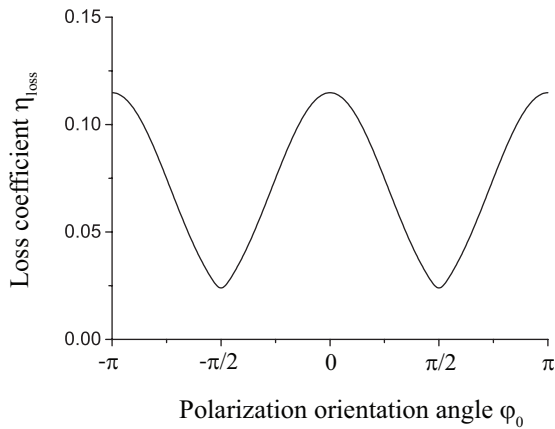}
 \end{center}
\caption{
Dependence of the loss coefficient $\eta_{\mathrm{loss}}$ on the azimuthal angle $\varphi_0$ of the field polarization vector $\mathbf{u}$ in the case where the guided probe field is quasilinearly polarized. 
The atom is located on the $x$ axis at the distance $r/a=1.8$.
Other parameters are as in Fig.~\ref{fig2}.
}
\label{fig10}
\end{figure}

It is not easy to measure the rate of scattering into forward guided modes directly in experiments. It is much more convenient to measure the loss of the field in forward guided modes. 
The loss rate $\Gamma_{\mathrm{loss}}$ is given by Eq.~\eqref{x10c}.
The lost power is $P_{\mathrm{loss}}=\hbar\omega \Gamma_{\mathrm{loss}}$. 
We introduce the loss coefficient $\eta_{\mathrm{loss}}=P_{\mathrm{loss}}/P_z$, which is related to the transmission $|T|^2$ as $\eta_{\mathrm{loss}}=1-|T|^2$.
We note that the loss coefficient $\eta_{\mathrm{loss}}$ can be considered as the generalized optical depth per atom for an array of atoms aligned in a line parallel to the fiber axis.
To get insight into scattering into forward guided modes, we plot in Fig.~\ref{fig9} the spatial and tuning dependencies of the loss coefficient $\eta_{\mathrm{loss}}$. In addition, we plot in Fig.~\ref{fig10} the dependence of the loss coefficient on the azimuthal  angle $\varphi_0$ of the polarization vector $\mathbf{u}$ of the guided probe field in the case where the latter is quasilinearly polarized.
The dashed red curve in Fig.~\ref{fig9}(a) shows that the maximal value of $\eta_{\mathrm{loss}}$, achieved for the guided light with the $x$ polarization and the atom at the distance $r/a=1$, is $\eta_{\mathrm{loss}}\simeq 0.59$.
Thus, in the vicinity of the fiber surface, the power $P_{\mathrm{loss}}$ lost by scattering into radiation modes and backward guided modes can be
up to 59\% of the propagation power $P_z$. This means that the transmittance of the guided probe field  
can be reduced to 41\%. It is clear from Fig.~\ref{fig9}(a) that the magnitude of the loss coefficient $\eta_{\mathrm{loss}}$ reduces with increasing radial distance $r$ and depends significantly on the polarization of the field. When the guided probe field is $y$-polarized, the maximal value of $\eta_{\mathrm{loss}}$, achieved at $r/a=1$, is $\eta_{\mathrm{loss}}\simeq 0.12$. When the guided probe field is quasicircularly polarized, the maximal value of $\eta_{\mathrm{loss}}$ is $\eta_{\mathrm{loss}}\simeq 0.40$. It is interesting that this value is far from the mean value between the values for the cases of the fields with the principal $x$ and $y$ polarizations. This deviation is a consequence of optical pumping in a multilevel atom. Due to the population redistribution, which is significant in the considered steady-state regime, the scattering efficiency of a quasicircularly polarized field is not the mean value of the scattering efficiencies of $x$ and $y$ polarized fields.
Figure \ref{fig10} shows that the loss coefficient $\eta_{\mathrm{loss}}$ is a symmetric and periodic function of $\varphi_0$ with the period of $\pi$.

The propagation power $P_z=10$ fW is small enough that any further decrease in $P_z$ would not practically change the scattering efficiency coefficients and the loss coefficient. 
Consequently, Figs.~\ref{fig2}--\ref{fig10} remain valid when $P_z\leq 10$ fW $=10^{-14}$ W.
Due to the population redistribution of the ground-state  sublevels,
the conventional result of the perturbation approach \cite{Boyd,cesium absorption}
for the scattering cross section in the steady-state regime cannot be used \cite{absorption} even though the atomic excitation is weak. We recognize that the results obtained for the atom on the fiber surface ($r/a=1$) are of only academic interest because the effect of the surface-induced potential is not considered in this paper. However, this effect can be neglected in the case of $r/a=1.8$ and $a=250$ nm, which corresponds to the situation realized in the experiment by Vetsch \textit{et al.} \cite{Vetsch10}.

\section{Directional spontaneous emission of a two-level atom}
\label{sec:twolevelatom}

In order to get deep insight into the asymmetry between the forward and backward scattering, 
we consider a two-level atom with a single upper level $|e\rangle$ and a single lower level $|g\rangle$ outside a nanofiber.
The rate of spontaneous emission into the guided modes with the positive ($f=+$) or negative ($f=-$) propagation direction is
\begin{equation}\label{t1}
\gamma^{(f)}=\gamma^{(fx)}+\gamma^{(fy)},
\end{equation}
where \cite{cesium decay} 
\begin{equation}\label{t2}
\gamma^{(f\xi)}=\frac{\omega_0\beta_0'}{2\epsilon_0\hbar}
\big|\mathbf{d}\cdot\mathbf{e}^{(\omega_0 f\xi)}\big|^2
\end{equation}
is the rate of spontaneous emission into the guided modes with the propagation direction $f=+,-$ and the polarization $\xi=x,y$.
We note that the dipole matrix-element vector $\mathbf{d}$ is, in general, a complex vector.
It is clear that $\gamma^{(f\xi)}$ and, consequently, $\gamma^{(f)}$ depend on the magnitude, the orientation, and the polarization of  
the dipole matrix-element vector $\mathbf{d}$. 
The rate $\Gamma_{f}$ of scattering into the guided modes with a given propagation direction $f$ is related to the corresponding spontaneous emission rate $\gamma^{(f)}$ as 
$\Gamma_{f}=\rho_{ee}\gamma^{(f)}$.

Assume that the atom is positioned on the positive side of the $x$ axis in the Cartesian coordinate system $\{x,y,z\}$, with $z$ being the fiber axis. In this case,  
we obtain the expressions 
\begin{equation}\label{t2c}
\mathbf{e}^{(\omega_0 fx)}=\sqrt2\;(i|e_r|,0,f|e_z|) 
\end{equation}
and
\begin{equation}\label{t2b}
\mathbf{e}^{(\omega_0 fy)}=\sqrt2\;(0,i|e_{\varphi}|,0) 
\end{equation}
for the profile functions of the $x$- and $y$-polarized guided modes, respectively [see Eqs.~\eqref{g9}]. 
It is clear from Eqs.~\eqref{t2} and \eqref{t2b} that the rate $\gamma^{(fy)}$ of spontaneous emission into the guided modes with the $y$ polarization does not depend on the emission direction $f$.
Meanwhile, it follows from Eqs.~\eqref{t2} and \eqref{t2c} that the rate $\gamma^{(fx)}$ of spontaneous emission into the guided modes with the $x$ polarization does not depend on the emission direction $f$ if the dipole components $d_z$ and $d_x$ have the same phase, that is,
if the atomic dipole is linearly polarized in the $zx$ plane.
In particular, the rates $\gamma^{(fx)}$ and $\gamma^{(f)}$ do not depend on $f$ in the cases 
where $\mathbf{d}$ is a real vector or the transition of the atom is of the type $\pi$ ($d_x=d_y=0$) 
or $\sigma_\pm$ ($d_z=0$ and $d_x=\pm id_y$) with respect to the fiber axis $z$.  However, the rates $\gamma^{(fx)}$ and $\gamma^{(f)}$ may depend on $f$ if both components $d_z$ and $d_x$ are nonzero and
have different phases, that is, if there is an ellipticity of the polarization of the atomic dipole matrix-element vector $\mathbf{d}$ in the $zx$ plane.
 
\begin{figure}[tbh]
\begin{center}
  \includegraphics{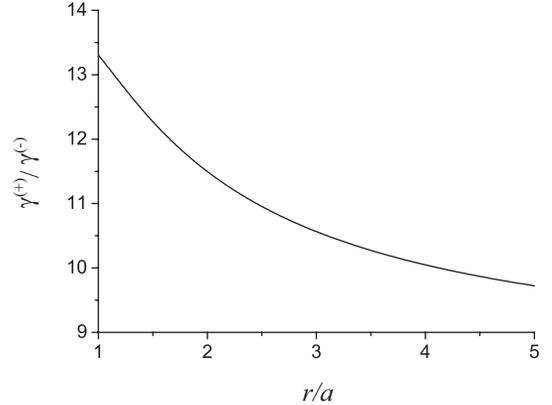}
 \end{center}
\caption{Ratio $\gamma^{(+)}/\gamma^{(-)}$ between the rates $\gamma^{(+)}$ and $\gamma^{(-)}$ of spontaneous emission from a two-level atom into the guided modes
in the directions $+\hat{\mathbf{z}}$ and $-\hat{\mathbf{z}}$, respectively. The levels of the atom are $|F'=5,M'=5\rangle$ and $|F=4,M=4\rangle$ of the $D_2$ line of atomic cesium 
with respect to the quantization axis $\hat{\mathbf{z}}_Q=\hat{\mathbf{y}}$.
The atom is positioned on the positive side of the $x$ axis. 
Other parameters are as in Fig.~\ref{fig2}.
}
\label{fig11}
\end{figure}

To illustrate such a situation, we consider the case where the atomic transition is of the type $\sigma_\pm$ with respect to the $y$ axis, that is, 
the polarization of the atomic dipole is circular in the $zx$ plane. In this case, it is natural to use the axis $y$ as the quantization axis.
The quantization coordinate system is $\{x_Q,y_Q,z_Q\}$, where
\begin{equation}\label{t2a}
x_Q=z, \qquad  y_Q=x, \qquad z_Q=y.
\end{equation} 
In this coordinate system, the dipole matrix element $\mathbf{d}$ of the atom has only a single nonzero spherical tensor component $d_q=-q(d_{x_Q}+iqd_{y_Q})/{\sqrt2}$, where $q=M_e-M_g=\pm1$ corresponds to
the transition type $\sigma_{\pm}$. An example of such a two-level atom is a cesium atom with the cycling transition
between the Zeeman levels $|F'=5, M'=\pm5\rangle$ and $|F=4, M=\pm4\rangle$ of the excited state $6P_{3/2}$ and the ground state $6S_{1/2}$, respectively.
The initial state can be prepared by optical pumping with the use of a circularly polarized field freely propagating along the $y$ direction.
In the Cartesian coordinate system $\{x,y,z\}$, the dipole matrix-element vector is
\begin{equation}\label{t3a}
\mathbf{d}=\frac{d_q}{\sqrt2}(i,0,-q).
\end{equation}
The above expression shows that, in the Cartesian coordinate system $\{x,y,z\}$, the dipole matrix-element vector $\mathbf{d}$ has two nonzero components, 
$d_z$ and $d_x=-iq d_z$, which are different in phase from each other by $\pi/2$. 
This is a consequence of the fact that the polarization of the atomic dipole is circular in the $zx$ plane. 
From Eq.~\eqref{t3a}, we find
\begin{equation}\label{t3}
\mathbf{d}\cdot\mathbf{e}^{(\omega_0 f\xi)}=\frac{d_q}{\sqrt2}(ie^{(\omega_0 f\xi)}_{x}-qe^{(\omega_0 f\xi)}_{z}).
\end{equation}
For the $y$-polarized guided modes, Eq.~\eqref{t2b} yields 
$e^{(\omega_0 fy)}_{x}=e^{(\omega_0 fy)}_{z}=0$. This leads to $\mathbf{d}\cdot\mathbf{e}^{(\omega_0 fy)}=0$ and, hence,
$\gamma^{(fy)}=0$. For the $x$-polarized guided modes, Eq.~\eqref{t2c} yields 
$e^{(\omega_0 fx)}_{x}=i\sqrt2 |e_r|$ and $e^{(\omega_0 fx)}_{z}= f\sqrt2|e_{z}|$.
This leads to 
\begin{equation}\label{t4}
\mathbf{d}\cdot\mathbf{e}^{(\omega_0 fx)}=-d_q(|e_{r}|+fq|e_{z}|).
\end{equation}
Hence, we find
\begin{eqnarray}\label{t5}
\gamma^{(f)}=\gamma^{(fx)}=\frac{\omega_0\beta'_0d_{q}^2}{2\epsilon_0\hbar}(|e_{r}|+fq|e_{z}|)^2.
\end{eqnarray}
It is clear that $\gamma^{(f)}$ depends on the emission direction $f=\pm$ and on the transition type $\sigma_\pm$ characterized by the number $q=M_e-M_g=\pm1$. Moreover, when we set $f=+$ and $f=-$ in Eq.~\eqref{t5} and then calculate the ratio between the results, we find
\begin{equation}\label{t6}
\frac{\gamma^{(+)}}{\gamma^{(-)}}\bigg|_{q=1}=\frac{\gamma^{(-)}}{\gamma^{(+)}}\bigg|_{q=-1} 
=\left(\frac{|e_{r}|+|e_{z}|}{|e_{r}|-|e_{z}|}\right)^{2}.
\end{equation}
Thus, the spontaneous emission of the two-level atom into the guided modes of the nanofiber may have different rates for different directions $f=\pm$.
We emphasize that the occurrence of $\gamma^{(+)}\not=\gamma^{(-)}$ is due to the existence of the longitudinal component $e_{z}$  
of the guided-mode profile function, the existence of the components $d_z$ and  $d_x$ of the atomic dipole matrix element,   
the ellipticity of the polarization of the $x$-polarized guided mode in the $zx$ plane, and the ellipticity of the polarization of the atomic dipole in the $zx$ plane. 
We note that the ratio $\gamma^{(+)}/\gamma^{(-)}$ is determined by just the ratio between the radial and axial components $e_r$ and $e_z$, respectively, of the guided-mode profile function. 
In Fig.~\ref{fig11}, we plot the dependence of the ratio $\gamma^{(+)}/\gamma^{(-)}$ on the radial position $r$ of the atom.
The figure shows that $\gamma^{(+)}/\gamma^{(-)}$ decreases slowly with increasing $r$ and can be as large as about $13.3$ for the atom on the fiber surface ($r/a=1$).  

We note that, in the case where the atom is positioned on the negative side of the $x$ axis,
we find from the first equation in Eqs.~\eqref{g9} the expressions
$e^{(\omega_0 fx)}_{x}|_{\varphi=\pi}=i\sqrt2 |e_r|$ and $e^{(\omega_0 fx)}_{z}|_{\varphi=\pi}=-f\sqrt2 |e_{z}|$ 
for the nonzero components of the profile function $\mathbf{e}^{(\omega_0 fx)}|_{\varphi=\pi}$ of the $x$-polarized guided modes. In this case, we obtain
\begin{equation}\label{t7}
\frac{\gamma^{(+)}}{\gamma^{(-)}}\bigg|_{q=1}=\frac{\gamma^{(-)}}{\gamma^{(+)}}\bigg|_{q=-1}
=\left(\frac{|e_{r}|-|e_{z}|}{|e_{r}|+|e_{z}|}\right)^{2}.
\end{equation}
It follows from the above results that the scattering of light from a system of two identical and equally excited two-level atoms, one above and one symmetrically below the fiber, into the guided modes has equal rates for the forward and backward directions.

According to the previous section, the asymmetry between the forward and backward scattering from an alkali-metal atom  into the guided modes is associated with the decay coefficients $\gamma_{e,e\pm1}^{(f)}$ and the atomic level coherences $\rho_{e\pm1,e}$ and therefore requires at least two upper levels. One may think that the explanation for the asymmetry in the case of a two-level atom contradicts that for the asymmetry in the case of a multilevel alkali-metal atom. However, the two pictures are consistent with each other. Indeed, the upper level of a two-level atom with a transition of the $\sigma_{\pm}$ type with respect to the $y$ axis corresponds to a superposition of several upper levels of a multilevel alkali-metal atom with the transitions of the $\pi$ and $\sigma_{\pm}$ types with respect to the $z$ axis.

\section{Summary}
\label{sec:summary}

We have presented a systematic theory for the scattering of guided light from a multilevel atom outside a nanofiber into the guided modes. In our numerical calculations, we used atomic cesium, with the transitions between the hyperfine levels $6S_{1/2}F=4$ and $6P_{3/2}F'=5$ of the $D_2$ line. We have calculated the scattering rate in the steady-state regime.
We have demonstrated analytically and numerically that the scattering rate is asymmetric with respect to the forward and backward directions and depends on the polarization of the probe field. The asymmetry between the forward and backward scattering is a result of the complexity of the level and transition structures of the atom and the existence of a longitudinal component of the guided-mode profile function. We have found that, in the case where the probe field is quasicircularly polarized, the rate of the scattering into the forward guided modes is a few times larger than that of the scattering into the backward guided modes. When the probe field is quasilinearly polarized along the axis $x$, which is the radial direction of the position of the atom, the rate of the scattering into the forward guided modes is about one order of magnitude larger than that of the scattering into the backward guided modes. However, when the probe field is quasilinearly polarized along the axis $y$, which is perpendicular to the radial direction of the position of the atom, the forward and backward scattering processes have the same rate. 

We have shown that, in the case of a two-level atom, the rates of spontaneous emission and consequently the rates
of scattering into the forward and backward guided modes differ from each other when the atomic dipole matrix-element vector
is a complex vector in the $zx$ plane, which contains the fiber axis $z$ and the atom-position radial axis $x$. 
In particular, for a two-level atom with the parameters of the cycling transition of atomic cesium, the ratio between the rates
of spontaneous emission (or scattering) into the forward and backward guided modes is as large as about $13.3$
for the atom on the surface of the nanofiber with the radius $a=250$ nm.
The directional spontaneous emission (or directional scattering) from such a two-level atom is a consequence
of the ellipticity of both the field polarization and the atomic dipole vector in the $zx$ plane.

\begin{acknowledgments}
We thank C. Clausen, C. Sayrin, and P. Schneeweiss for helpful comments and discussions.
F.L.K. acknowledges support by the Austrian Science Fund (Lise Meitner project No. M 1501-N27)
and by the European Commission (Marie Curie IIF Grant 332255). 
\end{acknowledgments}


\appendix

\section{Guided modes of a nanofiber}
\label{sec:guided}

Consider a nanofiber that is a silica cylinder of radius $a$ and refractive index $n_1$ and is surrounded by an infinite background medium of refractive index $n_2$,
where $n_2<n_1$. The radius of the nanofiber is well below a given free-space wavelength $\lambda$ of light. Therefore, the nanofiber supports only the hybrid fundamental modes HE$_{11}$ corresponding to the given wavelength $\lambda$ \cite{fiber books}. The light field in such a mode is strongly guided. It penetrates into the outside of the nanofiber in the form of an evanescent wave carrying a significant fraction of energy \cite{fibermode}.
For a fundamental guided mode HE$_{11}$ of a light field of frequency $\omega$ (free-space wavelength $\lambda=2\pi c/\omega$ and free-space wave number $k=\omega/c$), the propagation constant $\beta$ is determined by the
fiber eigenvalue equation \cite{fiber books}
\begin{eqnarray}\label{g1}
\frac{J_0(h a)}{h a J_1(h a)}&=&
-\frac{n_1^2+n_2^2}{2n_1^2}\frac{K_1'(q a)}{q a K_1(q a)}+ \frac{1}{h^2 a^2}
\nonumber\\&&\mbox{}
-\Bigg[\left(\frac{n_1^2-n_2^2}{2n_1^2}\frac{K_1'(q a)}{q a K_1(q a)}\right)^2
\nonumber\\&&\mbox{}
+\frac{\beta^2}{n_1^2 k^2}\left(\frac{1}{q^2a^2}+\frac{1}{h^2a^2}\right)^2 \Bigg]^{1/2}.
\end{eqnarray}
Here the parameters $h=(n_1^2k^2-\beta^2)^{1/2}$ and $q=(\beta^2-n_2^2k^2)^{1/2}$ characterize the fields inside and outside the fiber, respectively. The notations $J_n$ and $K_n$ stand for the Bessel functions of the first kind and the modified Bessel functions of the second kind, respectively. 

According to \cite{fiber books}, the cylindrical-coordinate vector components of the profile function $\mathbf{e}(\mathbf{r})$ 
of the electric part of the fundamental guided mode that propagates in the forward ($+\hat{\mathbf{z}}$) direction and is
counterclockwise quasicircularly polarized are given, for $r<a$, by
\begin{eqnarray}\label{g2}
e_{r}&=&iC\frac{q}{h}\frac{K_1(qa)}{J_1(ha)}[(1-s)J_0(hr)-(1+s)J_2(hr) ],
\nonumber\\
e_{\varphi}&=&-C\frac{q}{h}\frac{K_1(qa)}{J_1(ha)}[(1-s)J_0(hr)+(1+s)J_2(hr) ],
\nonumber\\
e_{z}&=&C\frac{2q}{\beta}\frac{K_1(qa)}{J_1(ha)}J_1(hr),
\end{eqnarray}
and, for $r>a$, by
\begin{eqnarray}\label{g3}
e_{r}&=&iC[(1-s)K_0(qr)+(1+s)K_2(qr) ],
\nonumber\\
e_{\varphi}&=&-C[(1-s)K_0(qr)-(1+s)K_2(qr) ],
\nonumber\\
e_{z}&=&C\frac{2q}{\beta}K_1(qr).
\end{eqnarray}
Here the parameter $s$ is defined as
\begin{equation}\label{g4} 
s=\frac{{1}/{h^2a^2}+{1}/{q^2a^2}}{{J_1^\prime (ha)}/{haJ_1(ha)}+{K_1^\prime (qa)}/{qaK_1(qa)}}.
\end{equation}
The parameter $C$ is the normalization coefficient. We take $C$ to be a positive real number and use the normalization condition 
\begin{equation}\label{g5}
\int _{0}^{2\pi}d\varphi\int _{0}^{\infty}n_{\mathrm{ref}}^2\,|\mathbf{e}|^2r\,dr=1.
\end{equation}
Here $n_{\mathrm{ref}}(r)=n_1$ for $r<a$, and $n_{\mathrm{ref}}(r)=n_2$ for $r>a$.
We note that the axial component $e_{z}$ is significant in the case of nanofibers \cite{fibermode}. This makes guided modes of nanofibers very different from plane-wave modes of the field in free space and from guided modes of conventional (weakly guiding) fibers \cite{fibermode,fiber books}.

We label quasicircularly polarized fundamental guided modes HE$_{11}$ by using a mode index $\mu=(\omega,f,l)$, where $\omega$ is the mode frequency, $f=+1$ or $-1$ (or simply $+$ or $-$) 
denotes the forward ($+\hat{\mathbf{z}}$) or backward ($-\hat{\mathbf{z}}$) propagation direction, respectively, and $l=+1$ or $-1$ (or simply $+$ or $-$) 
denotes the counterclockwise  or clockwise circulation, respectively, of the transverse component of the polarization around the axis $+\hat{\mathbf{z}}$. 
In the cylindrical coordinates, the components of the profile function $\mathbf{e}^{(\mu)}(\mathbf{r})$ of the electric part of the quasicircularly polarized fundamental guided mode $\mu$ are given by
\begin{eqnarray}\label{g6}
e_{r}^{(\mu)}&=&e_{r},
\nonumber\\
e_{\varphi}^{(\mu)}&=&le_{\varphi},
\nonumber\\
e_{z}^{(\mu)}&=& fe_{z}.
\end{eqnarray}
Consequently, the profile function of the quasicircularly polarized mode $(\omega, f, l)$ can be written as
\begin{eqnarray}\label{g7}
\mathbf{e}^{(\omega fl)}&=&\hat{\mathbf{r}}e^{(\omega fl)}_r+\hat{\boldsymbol{\varphi}}e^{(\omega fl)}_\varphi+\hat{\mathbf{z}}e^{(\omega fl)}_z
\nonumber\\ 
&=&\hat{\mathbf{r}}e_r+l\hat{\boldsymbol{\varphi}}e_\varphi+f\hat{\mathbf{z}}e_z,
\end{eqnarray}
where the notations 
\begin{eqnarray}\label{g6a}
\hat{\mathbf{r}} &=& \hat{\mathbf{x}}\cos\varphi + \hat{\mathbf{y}}\sin\varphi, \nonumber\\ 
\hat{\boldsymbol{\varphi}} &=& -\hat{\mathbf{x}}\sin\varphi + \hat{\mathbf{y}}\cos\varphi, 
\end{eqnarray}
and $\hat{\mathbf{z}}$ stand for the unit basis vectors of the cylindrical coordinate system $\{r,\varphi,z\}$.
Here $\hat{\mathbf{x}}$ and $\hat{\mathbf{y}}$ are the unit basis vectors of the Cartesian coordinate system for the fiber transverse plane $xy$.

We note that expression \eqref{g7} for the mode profile function $\mathbf{e}^{(\omega fl)}$ does not include the phase factor $e^{if\beta z +il\varphi}$,
which is present in the full expression for the electric part of the guided field in a quasicircularly polarized mode.
Indeed, the electric part $\boldsymbol{\mathcal{E}}_{\mathrm{circ}}^{(\omega fl)}$ of the guided field in the quasicircularly polarized mode $(\omega, f, l)$
is given by \cite{fiber books}
\begin{equation}\label{g7a}
\boldsymbol{\mathcal{E}}_{\mathrm{circ}}^{(\omega fl)} = A(\hat{\mathbf{r}}e_r+l\hat{\boldsymbol{\varphi}}e_\varphi+
f\hat{\mathbf{z}}e_z) e^{if\beta z +il\varphi},
\end{equation}
where the coefficient $A$ is determined by the power of the field.

Quasilinearly polarized guided modes are linear superpositions of quasicircularly polarized guided modes.
The electric part $\boldsymbol{\mathcal{E}}_{\mathrm{lin}}^{(\omega f\varphi_0)}$ of the guided field in a quasilinearly polarized mode $(\omega, f, \varphi_0)$
is given by \cite{fiber books}
\begin{equation}\label{g7b}
\boldsymbol{\mathcal{E}}_{\mathrm{lin}}^{(\omega f\varphi_0)} 
=\frac{1}{\sqrt2}( \boldsymbol{\mathcal{E}}_{\mathrm{circ}}^{(\omega f+)}e^{-i\varphi_0}+\boldsymbol{\mathcal{E}}_{\mathrm{circ}}^{(\omega f-)}e^{i\varphi_0}),
\end{equation}
where the angle $\varphi_0$ determines the orientation of the principal component of the polarization of the guided field in the fiber transverse plane.
In particular, the angle $\varphi_0=0$ or $\pi/2$ specifies the principal direction $x$ or $y$ of the polarization vector in the fiber transverse plane, respectively.
In terms of the components $e_r$, $e_\varphi$, and $e_z$ of the profile function of the quasicircularly polarized guided modes,  
we can rewrite $\boldsymbol{\mathcal{E}}_{\mathrm{lin}}^{(\omega f\varphi_0)}$ as
\begin{eqnarray}\label{7c}
\boldsymbol{\mathcal{E}}_{\mathrm{lin}}^{(\omega f\varphi_0)}
&=&\sqrt2 A[\hat{\mathbf{r}}e_r\cos(\varphi-\varphi_0)+i\hat{\boldsymbol{\varphi}}e_\varphi\sin(\varphi-\varphi_0)
\nonumber\\&&\mbox{}
+f\hat{\mathbf{z}}e_z\cos(\varphi-\varphi_0)] e^{if\beta z}.
\end{eqnarray}
Hence, we have 
\begin{equation}\label{g7d}
\boldsymbol{\mathcal{E}}_{\mathrm{lin}}^{(\omega f\varphi_0)} 
=A\mathbf{e}^{(\omega f\varphi_0)}e^{if\beta z},
\end{equation}
where
\begin{eqnarray}\label{g7e}
\mathbf{e}^{(\omega f\varphi_0)}&=&
\frac{1}{\sqrt2}(\mathbf{e}^{(\omega f+)}e^{i(\varphi-\varphi_0)}+\mathbf{e}^{(\omega f-)}e^{-i(\varphi-\varphi_0)})
\nonumber\\
&=&\sqrt2 [\hat{\mathbf{r}}e_r\cos(\varphi-\varphi_0)+i\hat{\boldsymbol{\varphi}}e_\varphi\sin(\varphi-\varphi_0)
\nonumber\\&&\mbox{}
+f\hat{\mathbf{z}}e_z\cos(\varphi-\varphi_0)]
\end{eqnarray}
is the profile function of the quasilinearly polarized guided mode $(\omega, f, \varphi_0)$.

In particular, the profile functions of quasilinearly polarized modes $(\omega,f,\xi)$, where $\xi=x$ or $y$, are given by
\begin{eqnarray}\label{g8}
\mathbf{e}^{(\omega fx)}&=&\frac{1}{\sqrt2}(\mathbf{e}^{(\omega f+)}e^{i\varphi}+\mathbf{e}^{(\omega f-)}e^{-i\varphi}),\nonumber\\
\mathbf{e}^{(\omega fy)}&=&\frac{1}{i\sqrt2}(\mathbf{e}^{(\omega f+)}e^{i\varphi}-\mathbf{e}^{(\omega f-)}e^{-i\varphi}).
\end{eqnarray}
In terms of the functions $e_r$, $e_\varphi$, and $e_z$, the profile functions $\mathbf{e}^{(\omega fx)}$ and $\mathbf{e}^{(\omega fy)}$ can be expressed as 
\begin{eqnarray}\label{g9}
\mathbf{e}^{(\omega fx)}&=&\sqrt2(\hat{\mathbf{r}}e_r\cos\varphi+i\hat{\boldsymbol{\varphi}}e_\varphi\sin\varphi+f\hat{\mathbf{z}}e_z\cos\varphi),\nonumber\\
\mathbf{e}^{(\omega fy)}&=&\sqrt2(\hat{\mathbf{r}}e_r\sin\varphi-i\hat{\boldsymbol{\varphi}}e_\varphi\cos\varphi+f\hat{\mathbf{z}}e_z\sin\varphi).\qquad
\end{eqnarray}

We introduce the notations $V_0=V_z$ and $V_{\pm 1}=\mp(V_x\pm i V_y)/\sqrt{2}$ for the spherical tensor components of an arbitrary vector $\mathbf{V}$. 
Due to the properties of the guided-mode profile functions \cite{fiber books}, we can represent
the spherical tensor components $e_{q}^{(\omega fl)}$ of the profile function $\mathbf{e}^{(\omega fl)}$ of the quasicircularly polarized guided mode $(\omega, f, l)$ in the form
\begin{equation}\label{g10}
e_{q}^{(\omega fl)}=f^{1+q}e^{iq(\varphi-\pi/2)}|e_{ql}|.
\end{equation}
Here we have introduced the notations
\begin{eqnarray}\label{g11} 
|e_0|&=&|e_z|,\nonumber\\
|e_{+1}|&=&\frac{|e_r|-|e_{\varphi}|}{\sqrt{2}},\nonumber\\ 
|e_{-1}|&=&\frac{|e_r|+|e_{\varphi}|}{\sqrt{2}}. 
\end{eqnarray}

We now examine the coefficients of spontaneous emission from a multilevel atom in the vicinity of a nanofiber into the guided modes. We use the notations $|e\rangle$ and $|g\rangle$ for the magnetic sublevels of the atom. According to Ref.~\cite{cesium decay}, the spontaneous emission from the atom into the guided modes of the nanofiber affects the evolution of the reduced density matrix of the atom through the set of decay coefficients 
\begin{eqnarray}\label{g12}
\gamma^{(\mathrm{gyd})}_{ee'gg'}&=&2\pi \sum_{f}\sum_{l=\pm}G_{\omega_0 fl eg}G_{\omega_0 fl e'g'}^* \nonumber\\
&=&2\pi \sum_{f}\sum_{\xi=x,y} G_{\omega_0 f\xi eg}G_{\omega_0 f\xi e'g'}^* , \nonumber\\
\gamma^{(\mathrm{gyd})}_{ee'}&=&2\pi \sum_{f}\sum_{l=\pm}\sum_{g}G_{\omega_0 fl eg}G_{\omega_0 fl e'g}^* \nonumber\\
&=&2\pi \sum_{f}\sum_{\xi=x,y}\sum_{g}G_{\omega_0 f\xi eg}G_{\omega_0 f\xi e'g}^*.
\end{eqnarray}
Here the coefficients 
\begin{eqnarray}\label{g13}
G_{\omega fl eg}=\sqrt{\frac{\omega\beta'}{4\pi\epsilon_0\hbar}}\;
\big(\mathbf{d}_{eg}\cdot\mathbf{e}^{(\omega fl)}\big)e^{i(f\beta z+l\varphi)}
\end{eqnarray}
with $l=+$ or $-$ characterize the coupling of the atomic transitions $|e\rangle\leftrightarrow |g\rangle$ with the quasicircularly polarized guided modes $(\omega, f, l)$, while the coefficients
\begin{eqnarray}\label{g14}
G_{\omega f\xi eg}=\sqrt{\frac{\omega\beta'}{4\pi\epsilon_0\hbar}}\;
\big(\mathbf{d}_{eg}\cdot\mathbf{e}^{(\omega f\xi)}\big)e^{if\beta z}
\end{eqnarray}
with $\xi=x$ or $y$ characterize the coupling of the atomic transitions $|e\rangle\leftrightarrow |g\rangle$ with the quasilinearly polarized guided modes $(\omega, f, \xi)$.
The notation $\beta'$ stands for the derivative of the propagation constant $\beta$ with respect to the frequency $\omega$.
The notation $\mathbf{d}_{eg}$ stands for the atomic dipole matrix element, which may be a complex vector.

We have the relations
\begin{eqnarray}\label{g15}
G_{\omega fx eg}&=&\frac{1}{\sqrt2}\big(G_{\omega f+ eg}+G_{\omega f- eg}\big),
\nonumber\\
G_{\omega fy eg}&=&\frac{1}{i\sqrt2}\big(G_{\omega f+ eg}-G_{\omega f- eg}\big),
\end{eqnarray}
which lead to
\begin{eqnarray}\label{g16}
\gamma_{ee'gg'}^{(fx)}&=&\frac{1}{2}\left(\gamma_{ee'gg'}^{(f+)}+\gamma_{ee'gg'}^{(f-)}+\gamma_{ee'gg'}^{(f+f-)}+\gamma_{ee'gg'}^{(f-f+)}\right),
\nonumber\\
\gamma_{ee'gg'}^{(fy)}&=&\frac{1}{2}\left(\gamma_{ee'gg'}^{(f+)}+\gamma_{ee'gg'}^{(f-)}-\gamma_{ee'gg'}^{(f+f-)}-\gamma_{ee'gg'}^{(f-f+)}\right),
\nonumber\\
\end{eqnarray}
and
\begin{eqnarray}\label{g17}
\gamma_{ee'}^{(fx)}&=&\frac{1}{2}\left(\gamma_{ee'}^{(f+)}+\gamma_{ee'}^{(f-)}+\gamma_{ee'}^{(f+f-)}+\gamma_{ee'}^{(f-f+)}\right),
\nonumber\\
\gamma_{ee'}^{(fy)}&=&\frac{1}{2}\left(\gamma_{ee'}^{(f+)}+\gamma_{ee'}^{(f-)}-\gamma_{ee'}^{(f+f-)}-\gamma_{ee'}^{(f-f+)}\right). \qquad 
\end{eqnarray}
Here we have introduced the notations
\begin{eqnarray}\label{g18}
\gamma_{ee'gg'}^{(fp)}&=&\gamma_{ee'gg'}^{(fp fp)},\nonumber\\
\gamma_{ee'}^{(fp)}&=&\gamma_{ee'}^{(fp fp)},
\end{eqnarray}
where
\begin{eqnarray}\label{g19}
\gamma_{ee'gg'}^{(fp f'p')}&=&2\pi G_{\omega_0fp eg}G_{\omega_0f'p' e'g'}^*,\nonumber\\
\gamma_{ee'}^{(fp f'p')}&=&\sum_g\gamma_{ee'gg}^{(fp f'p')}.
\end{eqnarray}
We introduce the notations
\begin{eqnarray}\label{g20}
\gamma_{ee'gg'}^{(f)}&=&\gamma_{ee'gg'}^{(fx)}+\gamma_{ee'gg'}^{(fy)}=\gamma_{ee'gg'}^{(f+)}+\gamma_{ee'gg'}^{(f-)},\nonumber\\
\gamma_{ee'}^{(f)}&=&\gamma_{ee'}^{(fx)}+\gamma_{ee'}^{(fy)}=\gamma_{ee'}^{(f+)}+\gamma_{ee'}^{(f-)}. 
\end{eqnarray}
We find the relations
\begin{eqnarray}\label{g21}
\gamma_{ee'gg'}^{(\mathrm{gyd})}&=&\gamma_{ee'gg'}^{(+)}+\gamma_{ee'gg'}^{(-)},\nonumber\\
\gamma_{ee'}^{(\mathrm{gyd})}&=&\gamma_{ee'}^{(+)}+\gamma_{ee'}^{(-)}.
\end{eqnarray}

According to Eq.~(\ref{x3}), only one spherical tensor component $d_{eg}^{(q)}\equiv (d_{eg})_q$ of the dipole vector $\mathbf{d}_{eg}$, with $q=M_e-M_g=-1$, 0, or 1, is nonzero. 
Hence, we obtain the formula
\begin{equation}\label{g22}
G_{\omega fl eg}=f^{1+q}e^{-iq\pi/2}e^{if\beta z}e^{i(l-q)\varphi}\sqrt{\frac{\omega\beta'}{4\pi\epsilon_0\hbar}}
d_{eg}^{(q)} |e_{-ql}|.
\end{equation}
For the opposite propagation directions $f$ and $\bar{f}=-f$, we find the relation
\begin{equation}\label{g23}
G_{\omega fl eg}=(-1)^{1+M_e-M_g}e^{2if\beta z}G_{\omega \bar{f}l eg},
\end{equation}
which yields
\begin{equation}\label{g24}
G_{\omega fl eg}G_{\omega fl' e'g'}^*=(-1)^{M_e-M_{e'}-M_g+M_{g'}}G_{\omega \bar{f}l eg}G_{\omega \bar{f}l' e'g'}^*.
\end{equation}
Hence, for the spontaneous emission coefficients $\gamma_{ee'}^{(fp fp')}$, given by Eq.~\eqref{x6}, we find the relation
\begin{equation}\label{g25}
\gamma_{ee'}^{(fp fp')}=(-1)^{M_e-M_{e'}}\gamma_{ee'}^{(\bar{f}p \bar{f}p')}. 
\end{equation} 
In particular, for the coefficients $\gamma_{ee'}^{(fp)}=\gamma_{ee'}^{(fpfp)}$, we obtain the relation
\begin{equation}\label{g26}
\gamma_{ee'}^{(fp)}=(-1)^{M_e-M_{e'}}\gamma_{ee'}^{(\bar{f}p)}.
\end{equation} 

We set $l=l'$ in Eq.~(\ref{g24}) and then apply the summations over $f$ and $l$. Then, we find the relation
\begin{equation}\label{g27}
\gamma^{(\mathrm{gyd})}_{ee'gg'}=(-1)^{M_e-M_{e'}-M_g+M_{g'}}\gamma^{(\mathrm{gyd})}_{ee'gg'},
\end{equation}
which yields
\begin{equation}\label{g28}
\gamma^{(\mathrm{gyd})}_{ee'}=(-1)^{M_e-M_{e'}}\gamma^{(\mathrm{gyd})}_{ee'}.
\end{equation}
It follows from Eq.~(\ref{g28}) that
\begin{equation}\label{g29}
\gamma^{(\mathrm{gyd})}_{e,e\pm1}=0.
\end{equation}

From Eq.~(\ref{g10}), we find
\begin{equation}\label{g30}
e_{q}^{(\omega fl)}=(-1)^q   e^{2iq\varphi} e_{-q}^{(\omega f\bar{l})},
\end{equation}
where $\bar{l}=-l$. On the other hand, when we use the properties of the Clebsch-Gordan coefficients and Eq.~(\ref{x3}), we find 
\begin{equation}\label{g31}
d_{eg}^{(q)}=(-1)^{F-F'+1} d_{\bar{e}\bar{g}}^{(\bar{q})},
\end{equation}
where $\bar{e}$ and $\bar{g}$ are the levels $|F',-M_e\rangle$ and $|F,-M_g\rangle$, respectively,
$q=M_e-M_g$, and $\bar{q}=-q$.
Then, we obtain the relation
\begin{equation}\label{g32}
G_{\omega fleg}=(-1)^{F-F'+1+M_e-M_g} e^{-2i(M_e-M_g-l)\varphi}G_{\omega f\bar{l}\bar{e}\bar{g}},
\end{equation}
which leads to
\begin{equation}\label{g33}
G_{\omega fleg}G_{\omega f'l'eg}^*=e^{2i(l-l')\varphi}G_{\omega f\bar{l}\bar{e}\bar{g}}G_{\omega f'\bar{l'}\bar{e}\bar{g}}^*,
\end{equation}
\begin{eqnarray}\label{g34}
G_{\omega fleg}G^*_{\omega fl'e'g}&=&(-1)^{M_e-M_{e'}} e^{-2i(M_e-M_{e'}-l+l')\varphi}\nonumber\\
&&\mbox{}\times
G_{\omega f\bar{l}\bar{e}\bar{g}}G^*_{\omega f\bar{l'}\bar{e'}\bar{g}},
\end{eqnarray}
and
\begin{eqnarray}\label{g35}
G_{\omega fleg}G^*_{\omega fle'g'}&=&e^{-i(M_e-M_{e'}-M_g+M_{g'})(2\varphi-\pi)}\nonumber\\
&&\mbox{}\times 
G_{\omega f\bar{l}\bar{e}\bar{g}}  G^*_{\omega f\bar{l}\bar{e'}\bar{g'}}.
\end{eqnarray}
When we apply the summation over $g$ to Eq.~(\ref{g34}), we find the relation
\begin{equation}\label{g36}
\gamma_{ee'}^{(flfl')}=(-1)^{M_e-M_{e'}} e^{-2i(M_e-M_{e'}-l+l')\varphi} \gamma_{\bar{e}\bar{e'}}^{(f\bar{l}f\bar{l'})},
\end{equation}
which yields 
\begin{equation}\label{g37}
\gamma_{ee'}^{(fl)}=e^{-i(M_e-M_{e'})(2\varphi-\pi )}\gamma_{\bar{e}\bar{e'}}^{(f\bar{l})}.
\end{equation}
When we take into account the relation (\ref{g26}), we obtain
\begin{equation}\label{g38}
\gamma_{ee'}^{(fl)}=e^{-2i(M_e-M_{e'})\varphi}\gamma_{\bar{e}\bar{e'}}^{(\bar{f}\bar{l})}.
\end{equation}
We apply the summations over $f$ and $l$ to Eq.~(\ref{g35}) and use the property (\ref{g27}) to simplify the result. Then, we obtain the relation 
\begin{equation}\label{g39}
\gamma_{ee'gg'}^{(\mathrm{gyd})}=e^{-2i(M_e-M_{e'}-M_g+M_{g'})\varphi}
\gamma_{\bar{e}\bar{e'}\bar{g}\bar{g'}}^{(\mathrm{gyd})},
\end{equation}
which leads to
\begin{equation}\label{g40}
\gamma_{ee'}^{(\mathrm{gyd})}=e^{-2i(M_e-M_{e'})\varphi}\gamma_{\bar{e}\bar{e'}}^{(\mathrm{gyd})}.
\end{equation}

It follows from Eq.~\eqref{g22} that the complex number $G_{\omega fl eg} G^*_{\omega fl' e'g}$ and its complex conjugate are related to each other as
\begin{eqnarray}\label{g41}
\lefteqn{G_{\omega fl eg} G^*_{\omega fl' e'g'}=(-1)^{M_e-M_{e'}-M_g+M_{g'}}}\nonumber\\
&&\mbox{}\times
e^{2i(l-l'-M_e+M_{e'}+M_g-M_{g'})\varphi}G^*_{\omega fl eg} G_{\omega fl' e'g'}. \quad
\end{eqnarray}
When we set $l=l'$ in Eq.~(\ref{g41}), apply the summations over $f$ and $l$, and use the property \eqref{g27} to simplify the result,
we obtain the relation 
\begin{equation}\label{g42}
\gamma_{ee'gg'}^{(\mathrm{gyd})}=e^{-2i(M_e-M_{e'}-M_g+M_{g'})\varphi} \gamma^{(\mathrm{gyd})*}_{ee'gg'},
\end{equation}
which yields
\begin{equation}\label{g43}
\gamma_{ee'}^{(\mathrm{gyd})}=e^{-2i(M_e-M_{e'})\varphi} \gamma^{(\mathrm{gyd})*}_{ee'}.
\end{equation}
When we set $g=g'$ in Eq.~(\ref{g41}) and apply the summation over $g$, we get the relation
\begin{equation}\label{g44}
\gamma_{ee'}^{(flfl')}=(-1)^{M_e-M_{e'}} e^{2i(l-l'-M_e+M_{e'})\varphi} \gamma^{(flfl')*}_{ee'},  
\end{equation}
which leads to
\begin{equation}\label{g45}
\gamma_{ee'}^{(fl)}=e^{-i(M_e-M_{e'})(2\varphi-\pi)} \gamma^{(fl)*}_{ee'}.  
\end{equation}

\section{Radiation modes of a nanofiber}
\label{sec:radiation}

For the radiation modes, we have $-kn_2<\beta<kn_2$.
The characteristic parameters for the field in the inside and outside of the fiber are $h=\sqrt{k^2n_1^2-\beta^2}$ and $q=\sqrt{k^2n_2^2-\beta^2}$, respectively.
The mode functions of the electric parts of the radiation modes $\nu=(\omega\beta m l)$
\cite{fiber books} are given, for $r<a$, by
\begin{eqnarray}\label{q1}
e_r^{(\nu)}&=&
\frac{i}{h^2}\left[\beta hAJ'_m(hr)+im\frac{\omega\mu_0}{r}BJ_m(hr)\right],\nonumber\\ 
e_{\varphi}^{(\nu)}&=&
\frac{i}{h^2}\left[im\frac{\beta}{r}AJ_m(hr)-h\omega\mu_0BJ'_m(hr)\right],\nonumber\\
e_z^{(\nu)}&=&AJ_m(hr),
\end{eqnarray}
and, for $r>a$, by 
\begin{eqnarray}\label{q2}
e_r^{(\nu)}&=&
\frac{i}{q^2}\sum_{j=1,2}
\left[\beta q C_jH^{(j)\prime}_m(qr)+im\frac{\omega\mu_0}{r}D_jH^{(j)}_m(qr)\right],\nonumber\\
e_{\varphi}^{(\nu)}&=&
\frac{i}{q^2}\sum_{j=1,2}
\left[im\frac{\beta}{r}C_jH^{(j)}_m(qr)-q\omega\mu_0D_jH^{(j)\prime}_m(qr)\right], \nonumber\\
e_z^{(\nu)}&=&\sum_{j=1,2}C_jH_m^{(j)}(qr).
\end{eqnarray}
Here $A$ and $B$ as well as $C_j$ and $D_j$ with $j=1,2$ are coefficients.
The coefficients $C_j$ and $D_j$ are related to the coefficients $A$ and $B$ as 
\cite{Tromborg}
\begin{eqnarray}\label{q3}
C_j&=&(-1)^{j}\frac{i\pi q^2a}{4n_2^2}(AL_j+i\mu_0cBV_j),\nonumber\\
D_j&=&(-1)^{j-1}\frac{i\pi q^2a}{4}(i\epsilon_0cAV_j-BM_j),
\end{eqnarray}
where
\begin{eqnarray}\label{q4}
V_j&=&\frac{mk\beta}{ah^2q^2}
(n_2^2-n_1^2)
J_m(ha)H_m^{(j)*}(qa),\nonumber\\
M_j&=&\frac{1}{h}J'_m(ha)H_m^{(j)*}(qa)
-\frac{1}{q}J_m(ha)H_m^{(j)*\prime}(qa),\nonumber\\
L_j&=&\frac{n_1^2}{h}J'_m(ha)H_m^{(j)*}(qa)
-\frac{n_2^2}{q}J_m(ha)H_m^{(j)*\prime}(qa).\nonumber\\
\end{eqnarray}
We specify two polarizations by choosing $B=i\eta A$ and $B=-i\eta A$ for $l=+$
and $l=-$, respectively. We take $A$ to be a real number
The orthogonality of the modes requires
\begin{eqnarray}\label{q5}
&&\int _0^{2\pi}d\varphi\int _{0}^{\infty}n_{\mathrm{ref}}^2
\left[\mathbf{e}^{(\nu)}\mathbf{e}^{(\nu')*}\right]_{\beta=\beta',m=m'}
\;rdr \nonumber\\&&
=N_{\nu}\delta_{ll'}\delta(\omega-\omega').
\end{eqnarray}
This leads to
\begin{equation}\label{q6}
\eta=\epsilon_0c\sqrt{\frac{n_2^2|V_j|^2+|L_j|^2}{|V_j|^2+n_2^2|M_j|^2}}.
\end{equation}
The constant $N_{\nu}$ is given by 
\begin{equation}\label{q7}
N_{\nu}=\frac{8\pi \omega}{q^2}\left(n_2^2|C_j|^2+\frac{\mu_0}{\epsilon_0}|D_j|^2\right).
\end{equation}
We use the normalization $N_{\nu}=1$.

We have the following symmetry relations: 
\begin{eqnarray}\label{q8}
e_r^{(\omega,\beta, m,l)}&=&-e_r^{(\omega,-\beta, m,-l)},\nonumber\\
e_{\varphi}^{(\omega,\beta, m,l)}&=&-e_{\varphi}^{(\omega,-\beta, m,-l)},\nonumber\\
e_z^{(\omega,\beta, m,l)}&=&e_z^{(\omega,-\beta, m,-l)},\nonumber\\
\end{eqnarray}
\begin{eqnarray}\label{q9}
e_{r}^{(\omega,\beta, m,l)}&=&(-1)^m e_{r}^{(\omega,\beta, -m,-l)},\nonumber\\
e_{\varphi}^{(\omega,\beta, m,l)}&=&(-1)^{m+1} e_{\varphi}^{(\omega,\beta, -m,-l)},\nonumber\\
e_{z}^{(\omega,\beta, m,l)}&=&(-1)^m e_{z}^{(\omega,\beta, -m,-l)},
\end{eqnarray} 
and
\begin{equation}\label{q10}
e_r^{(\nu)*}=-e_r^{(\nu)},\quad
e_\varphi^{(\nu)*}=e_\varphi^{(\nu)},\quad
e_z^{(\nu)*}=e_z^{(\nu)}.
\end{equation}
For the spherical tensor components $e_q^{(\omega,\beta, m,l)}$, with the index $q=0,\pm1$, of the radiation mode functions, we find the relations
\begin{equation}\label{q11}
e_q^{(\omega\beta m l)}=(-1)^q e_q^{(\omega\bar{\beta} m \bar{l})},
\end{equation}
\begin{equation}\label{q12}
e_{q}^{(\omega\beta m l)}=(-1)^{m+q} e^{2iq\varphi} e_{-q}^{(\omega\beta \bar{m} \bar{l})},
\end{equation}
and
\begin{equation}\label{q13}
e_{q}^{(\omega\beta m l)}=(-1)^{q} e^{2iq\varphi} e_{q}^{(\omega\beta m l)*}.
\end{equation}
Here we have introduced the notations $\bar{\beta}=-\beta$, $\bar{m}=-m$, and $\bar{l}=-l$.
 
We now examine the coefficients of spontaneous emission from a multilevel atom in the vicinity of a nanofiber into the radiation modes. 
We use the notations $|e\rangle$ and $|g\rangle$ for the magnetic sublevels of a multilevel atom in the vicinity of the nanofiber.
According to Ref.~\cite{cesium decay}, the spontaneous emission from the atom into the radiation modes of the nanofiber affects the evolution of the atomic reduced density matrix through the set of decay coefficients 
 \begin{eqnarray}\label{q14}
\gamma^{(\mathrm{rad})}_{ee'gg'}&=&2\pi \sum_{ml}\int_{-k_0n_2}^{k_0n_2}d\beta\,G_{\nu_0 eg}G_{\nu_0 e'g'}^*,
\nonumber\\
\gamma^{(\mathrm{rad})}_{ee'}&=&2\pi \sum_{mlg}\int_{-k_0n_2}^{k_0n_2}d\beta\, G_{\nu_0 eg}G_{\nu_0 e'g}^*.
\end{eqnarray}
Here $\nu_0=(\omega_0,\beta,m,l)$ labels resonant radiation modes and
\begin{equation}\label{q15}
G_{\nu eg}=\sqrt{\frac{\omega}{4\pi\epsilon_0\hbar}}\;
\big(\mathbf{d}_{eg}\cdot\mathbf{e}^{(\nu)}\big)e^{i(\beta z+m\varphi)}
\end{equation}
characterizes the coupling of the atomic transition $|e\rangle\leftrightarrow |g\rangle$ with
the radiation mode $\nu=(\omega,\beta,m,l)$. 

It follows from the property (\ref{q11}) that the coupling coefficients 
$G_{\nu eg}=G_{\omega\beta m l eg}$ satisfy the relation
\begin{equation}\label{q16}
G_{\omega\beta m l eg}=(-1)^{M_e-M_g}e^{2i\beta z} G_{\omega\bar{\beta} m \bar{l} eg},
\end{equation}
which leads to
\begin{eqnarray}\label{q17}
\lefteqn{G_{\omega\beta m l eg} G^*_{\omega\beta m l e'g'}}\nonumber\\
&&=(-1)^{M_e-M_{e'}-M_g+M_{g'}} G_{\omega\bar{\beta} m \bar{l} eg} G^*_{\omega\bar{\beta} m \bar{l} e'g'}.
\end{eqnarray}
When we apply the integration over $\beta$ and the summations over $m$ and $l$ to Eq.~(\ref{q17}), we find the relation
\begin{equation}\label{q18}
\gamma^{(\mathrm{rad})}_{ee'gg'}=(-1)^{M_e-M_{e'}-M_g+M_{g'}}\gamma^{(\mathrm{rad})}_{ee'gg'},
\end{equation}
which yields
\begin{equation}\label{q19}
\gamma^{(\mathrm{rad})}_{ee'}=(-1)^{M_e-M_{e'}}\gamma^{(\mathrm{rad})}_{ee'}.
\end{equation}
It follows from Eq.~(\ref{q19}) that
\begin{equation}\label{q20}
\gamma^{(\mathrm{rad})}_{e,e\pm1}=0.
\end{equation}

When we use Eqs.~(\ref{g31}) and (\ref{q12}), we find from expression (\ref{q15}) the relation
\begin{equation}\label{q21}
G_{\omega\beta m l eg}=(-1)^{F-F'+1} e^{-i(M_e-M_g-m)(2\varphi-\pi)}G_{\omega\beta \bar{m} \bar{l} \bar{e}\bar{g}},
\end{equation}
which leads to
\begin{eqnarray}\label{q22}
G_{\omega\beta m l eg} G^*_{\omega\beta m l e'g'}&=&e^{-i(M_e-M_{e'}-M_g+M_{g'})(2\varphi-\pi)}\nonumber\\
&&\mbox{}\times 
G_{\omega\beta \bar{m} \bar{l} \bar{e}\bar{g}} G^*_{\omega\beta \bar{m} \bar{l} \bar{e'}\bar{g'}}.
\end{eqnarray}
We apply the integration over $\beta$ and the summations over $m$ and $l$ to Eq.~(\ref{q22}) 
and use the property (\ref{q18}) to simplify the result. Then, we find the relation
\begin{equation}\label{q23}
\gamma_{ee'gg'}^{(\mathrm{rad})}=e^{-2i(M_e-M_{e'}-M_g+M_{g'})\varphi}
\gamma_{\bar{e}\bar{e'}\bar{g}\bar{g'}}^{(\mathrm{rad})}
\end{equation}
and, hence,
\begin{equation}\label{q24}
\gamma_{ee'}^{(\mathrm{rad})}=e^{-2i(M_e-M_{e'})\varphi}\gamma_{\bar{e}\bar{e'}}^{(\mathrm{rad})}.
\end{equation}

It follows from Eq.~\eqref{q13} and expression \eqref{q15} that the coupling coefficient $G_{\nu eg}$ and its complex conjugate are related to each other as
\begin{equation}\label{q25}
G_{\nu eg}=e^{-iq(2\varphi-\pi)}e^{2i(\beta z+m\varphi)}G_{\nu eg}^*.
\end{equation}
Hence, we have
\begin{equation}\label{q26}
G_{\nu eg}G_{\nu e'g'}^*=e^{-i(M_e-M_{e'}-M_g+M_{g'})(2\varphi-\pi)}
G_{\nu eg}^* G_{\nu e'g'}.
\end{equation}
We apply the summation over $\nu$ (i.e., the integration over $\beta$ and the summations over $m$ and $l$) to Eq.~(\ref{q26}) 
and use the property (\ref{q18}) to simplify the result. Then, we find the relation
\begin{equation}\label{q27}
\gamma_{ee'gg'}^{(\mathrm{rad})}=e^{-2i(M_e-M_{e'}-M_g+M_{g'})\varphi}
\gamma_{ee'gg'}^{(\mathrm{rad})*}.
\end{equation}

\end{document}